\documentclass[journal]{IEEEtran}

\usepackage{cite}
\usepackage{float}

\ifCLASSINFOpdf
\usepackage[pdftex]{graphicx}
\usepackage{subcaption}
 \DeclareGraphicsExtensions{.eps}

\usepackage{amsmath}
\usepackage{amssymb}
\usepackage{bm}
\usepackage{diagbox}

\usepackage{algorithmic}
\usepackage{lipsum}
\usepackage{footnote}
\usepackage{hyperref}

\usepackage{bbm}

\usepackage{url}
\usepackage{threeparttable}
\usepackage{multirow}

\usepackage{color}
\begin{document}
\bibliographystyle{IEEEtran}
\title{MethConvTransformer: A Deep Learning Framework for Cross-Tissue Alzheimer’s Disease Detection}

\author{Gang~Qu, Guanghao~Li, Zhongming~Zhao, for the Alzheimer's Disease Neuroimaging Initiative*% <-this % stops a space

\IEEEcompsocitemizethanks{
\IEEEcompsocthanksitem G.Qu, G.Li, and Z.Zhao are with McWilliams School of Biomedical Informatics and School of Public Health, The University of Texas Health Science Center at Houston, Houston, TX, 77030.\protect\\
Corresponding author: Zhongming Zhao.  Zhongming.Zhao@uth.tmc.edu\\
*Data used in preparation of this article were obtained from the Alzheimer's Disease
Neuroimaging Initiative (ADNI) database (adni.loni.usc.edu). As such, the investigators within the ADNI contributed to the design 
and implementation of ADNI and/or provided data but did not participate in analysis or writing of this report. 
A complete listing of ADNI investigators can be found at:
\url{http://adni.loni.usc.edu/wp-content/uploads/how_to_apply/ADNI_Acknowledgement_List.pdf}\protect
}% <-this % stops an unwanted space
}

\maketitle
\begin{abstract}
Alzheimer’s disease (AD) is a multifactorial neurodegenerative disorder characterized by progressive cognitive decline and widespread epigenetic dysregulation in the brain. DNA methylation, as a stable yet dynamic epigenetic modification, holds promise as a noninvasive biomarker for early AD detection. However, methylation signatures vary substantially across tissues and studies, limiting reproducibility and translational utility. To address these challenges, we develop \textbf{MethConvTransformer}, a transformer-based deep learning framework that integrates DNA methylation profiles from both brain and peripheral tissues to enable biomarker discovery. The model couples a CpG-wise linear projection with convolutional and self-attention layers to capture local and long-range dependencies among CpG sites, while incorporating subject-level covariates and tissue embeddings to disentangle shared and region-specific methylation effects. In experiments across six GEO datasets and an independent ADNI validation cohort, our model consistently outperforms conventional machine-learning baselines, achieving superior discrimination and generalization. Moreover, interpretability analyses using linear projection, SHAP, and Grad-CAM++ reveal biologically meaningful methylation patterns aligned with AD-associated pathways, including immune receptor signaling, glycosylation, lipid metabolism, and endomembrane (ER/Golgi) organization. Together, these results indicate that MethConvTransformer delivers robust, cross-tissue epigenetic biomarkers for AD while providing multi-resolution interpretability, thereby advancing reproducible methylation-based diagnostics and offering testable hypotheses on disease mechanisms.
\end{abstract}

\IEEEpeerreviewmaketitle

\section{Introduction}\label{sec:introduction}
Alzheimer's disease (AD) is the most common form of dementia and represents an escalating public health challenge. In the United States, an estimated 7.2 million adults aged 65 years and older are currently living with AD, and this number is projected to nearly double to 13.8 million by 2060 in the absence of effective preventive or curative interventions according to the Alzheimer’s Association 2025 report \cite{ADfact2025}. AD is a leading cause of death, with over 120,000 deaths recorded in 2022, making AD the seventh leading cause of death overall in the United States \cite{xu2025deaths}. The progressive cognitive decline associated with AD not only places a substantial burden on patients but also on family caregivers and the health care system. Given these challenges, there is an urgent need for sensitive and accessible biomarkers that can detect AD in its prodromal stage. 

Epigenetic markers, such as DNA methylation, are increasingly recognized as promising candidates. DNA methylation typically occurs at cytosine residues within CpG dinucleotides, where a methyl group is covalently added to the 5' position of cytosine to form 5-methylcytosine (5mC). This modification can modulate gene expression by recruiting repressive proteins or by blocking transcription factor binding \cite{moore2013dna}. During development, cells acquire lineage- and tissue-specific methylation patterns \cite{horvath2013dna} that are sufficiently stable to serve as molecular signatures of tissue identity and disease processes \cite{sarnowski2023multi}. These properties have made DNA methylation an attractive candidate biomarker for neurodegenerative disorders including AD.

Previous work has revealed that AD is associated with methylation changes in the brain \cite{lunnon2014methylomic, de2014alzheimer}, but the magnitude and consistency of these changes vary substantially across brain regions and studies \cite{ yokoyama2017dna, lang2022methylation}. In the context of AD, large-scale epigenome--wide association studies (EWAS) have evealed robust methylation differences in cortical tissue, many of which map to genes implicated in immune regulation, synaptic function, and glial responses\cite{zhang2020epigenome, shireby2022dna, gasparoni2018dna}, thereby aligning with known pathological processes in AD. By contrast, findings from blood-derived methylation data have been less consistent: although some studies report correlations between blood methylation and cerebrospinal fluid biomarkers \cite{smith2024blood}, replication across independent cohorts has remained challenging, likely due to differences in tissue specificity, cohort composition, and analytic methodology \cite{fransquet2018blood, kaleck2025replication}. This tissue-specific discrepancy highlights a translational dilemma for biomarker development, as brain-derived methylation profiles are closely related to disease pathology but obtainable only postmortem or through invasive procedures, whereas blood is more accessible for clinical applications but may yield weaker or less reproducible signatures.

Given these tissue-specific discrepancies, computational models have been developed to improve the identification of methylation-based biomarkers. Traditional machine learning classifiers, such as support vector machines and logistic regression, typically rely on selected CpG subsets or dimensionality-reduced features, but they are constrained to linear decision rules and struggle to capture higher-order, non-linear dependencies across hundreds of thousands of loci \cite{ren2020identification}. In the realm of brain methylation, EWASplus successfully discriminated AD cases from controls using CpG features derived from cortical EWAS data \cite{huang2021machine}. Recent advances in deep learning have been applied to longitudinal methylation data to predict Alzheimer's progression, showing improved performance in modeling complex patterns \cite{chen2022multi}. However, these models are usually developed on single-tissue datasets (e.g., brain or blood) and do not leverage the complementary information available across tissues, which constrains their translational potential. To address this limitation, Silva et al. \cite{c2022cross} explored cross-tissue integration of blood and brain methylation and identified AD-associated CpGs; however, the number of consistent signals was modest and predictive performance remained limited. Because current cross-tissue strategies are largely based on meta-analysis or regression, they fail to adequately model heterogeneous and tissue-specific regulatory mechanisms. This underscores the need for novel integrative frameworks capable of capturing complex cross-tissue relationships to improve biomarker discovery and mechanistic insight in AD.

\begin{figure*}[htp!]
  \centering
  \includegraphics[width=0.9\textwidth]{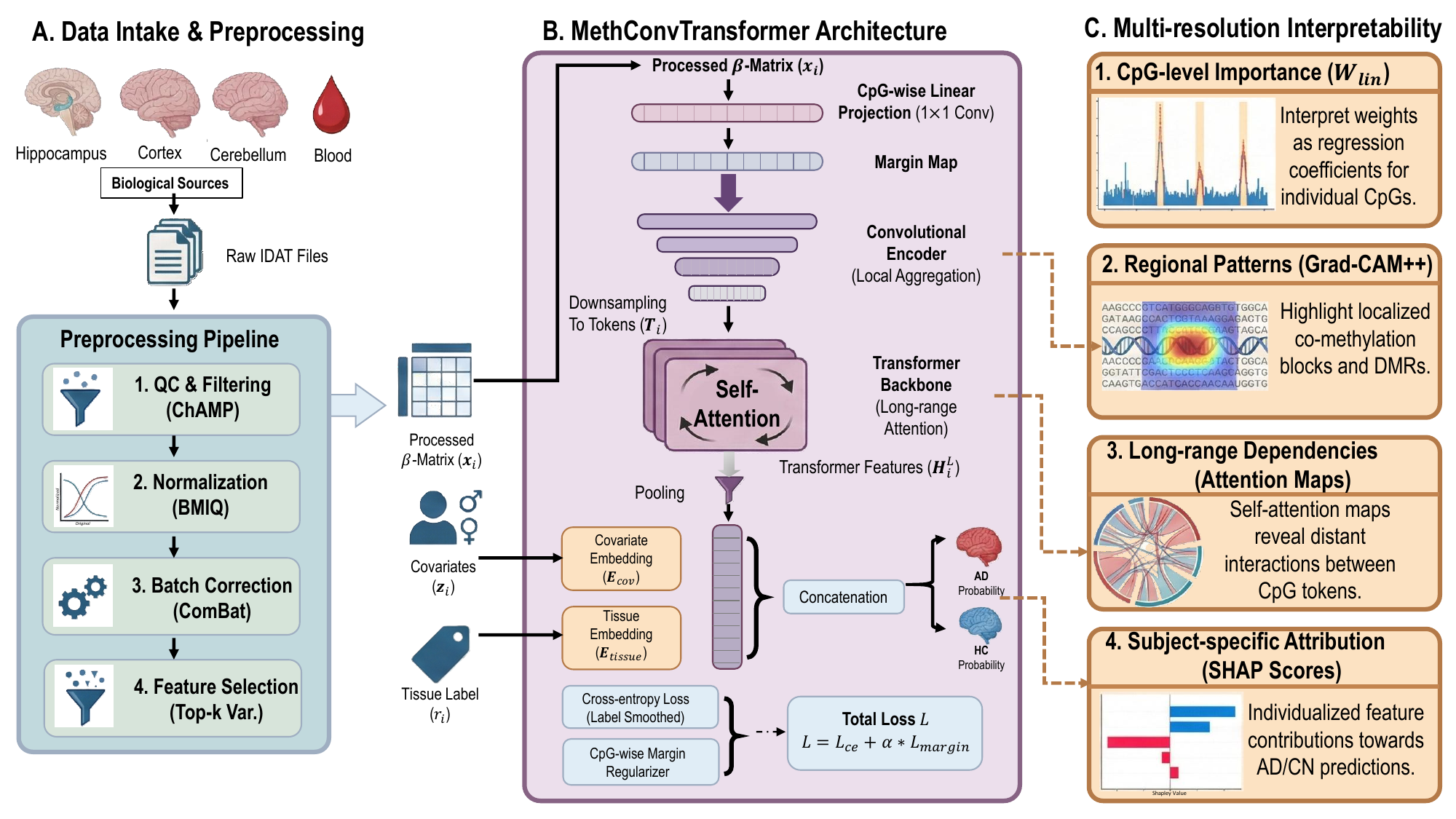}
  \caption{Overview of the MethConvTransformer framework for DNA methylation analysis:
(A) Preprocessing pipeline converting raw IDAT files from multiple brain regions and blood into normalized, batch-corrected $\beta$-matrices with selected CpG features; (B) MethConvTransformer architecture combining CpG-wise linear projection, local convolutional encoding, and Transformer-based long-range attention, jointly integrated with covariate and tissue embeddings for AD prediction; (C) Multi-resolution interpretability module capturing CpG-level importance, regional co-methylation patterns (Grad-CAM++), long-range dependencies (attention maps), and subject-specific feature attribution (SHAP).}
  \label{fig:pipeline}
\end{figure*}

Building on these limitations, we propose a multi-tissue epigenetic integration framework for AD detection. The model leverages a transformer-based \cite{qu2023interpretable} architecture capable of capturing long-range and non-linear dependencies among CpG sites. It incorporates covariates such as age and sex, applies index-based positional encodings to represent CpG sites, and learns tissue embeddings to disentangle shared versus tissue-specific methylation patterns. By training on heterogeneous cohorts spanning brain and peripheral tissues, our framework is designed to improve the sensitivity and generalizability of methylation-based diagnostics, while offering mechanistic insights into the cross-tissue epigenetic processes underlying AD. Importantly, our framework integrates an interpretability module that quantifies the contribution of individual CpG sites to model predictions and enables a unified assessment of feature importance across tissues. This capability goes beyond conventional differential methylation analyses, which typically rely on statistical significance thresholds at the single-site level. Whereas differential analysis can identify robust site-specific changes, it often overlooks weak or combinatorial effects that may be biologically meaningful. Our approach complements and extends such analyses by capturing higher-order, cross-tissue patterns and ranking CpG importance in a continuous and interpretable manner, thereby bridging predictive modeling with mechanistic insight.


\section{Materials and Methods}
\label{method}
\subsection{Data acquisition and cohorts}
The study will utilise publicly available DNA methylation datasets generated with Illumina 450K or EPIC arrays.  Data from multiple tissues (e.g., brain regions and peripheral blood) will be collected to facilitate cross--tissue modelling. Specifically, we compiled multiple publicly available DNA methylation datasets from the Gene Expression Omnibus (GEO) repository across both brain and blood tissues. Table~\ref{tab:datasets} summarizes the datasets included in this study, their tissue regions, sample sizes, platforms, and available data formats. Across cohorts, we retrieved publicly available DNAm datasets from GEO generated on Illumina 450K or EPIC arrays. Inclusion required availability of raw IDAT pairs (or equivalent probe-level intensities) and core sample metadata sufficient to define tissue/region and disease status. We targeted brain regions and peripheral blood; datasets lacking unambiguous tissue annotation or harmonizable labels were excluded. When duplicate assays or technical replicates existed for the same subject–region, all records were retained at acquisition (no de-duplication applied in this iteration). 

\begin{table*}[htbp!]
\centering
\caption{Summary of DNA methylation datasets included in this study.}
\label{tab:datasets}
\begin{tabular}{p{2.2cm} p{3.5cm} p{1.5cm} p{2.2cm} p{2.2cm} p{2.8cm}}
\hline
\textbf{Dataset} & \textbf{Tissue / Region} & \textbf{Samples} & \textbf{Platform} & \textbf{Data Format} & \textbf{Sample Groups} \\
\hline
GSE125895\cite{semick2019integrated} & Entorhinal cortex (ERC); Hippocampus; DLPFC; Cerebellum & 269  & Illumina 450K & IDAT & 82 AD, 187 CN \\
GSE134379\cite{brokaw2020cell} & Middle temporal gyrus; Cerebellum & 404 & Illumina 450K & IDAT & 225 AD, 179 CN \\
GSE66351\cite{gasparoni2018dna} & Frontal cortex neurons \& glia  & 192 & Illumina 450K & IDAT & 106 AD, 84 CN \\
GSE59685\cite{lunnon2014methylomic}& ERC; Superior temporal gyrus; Prefrontal cortex; Cerebellum; Whole blood & 531 & Illumina 450K & $\beta$-matrix & 207 AD, 103 CN\\
GSE80970\cite{smith2018elevated}& Prefrontal cortex; Superior temporal gyrus & 286 & Illumina 450K & $\beta$-matrix & 148 AD, 138 CN \\
GSE144858\cite{roubroeks2020epigenome} & Whole blood & 300 & Illumina 450K & $\beta$-matrix & 93 AD, 111 MCI, 96 CN \\
\hline
\end{tabular}
\end{table*}

All datasets were retrieved from GEO and screened to include subjects with clear diagnostic labels (AD, MCI, or cognitively normal controls). For multi-region studies, only samples with complete metadata were retained. We excluded:  
\begin{itemize}
    \item samples lacking age, sex, or diagnosis information,  
    \item technical replicates or low-quality arrays flagged in the original GEO records,  
    \item samples with missing or corrupted IDAT/$\beta$-matrix files.  
\end{itemize}

We assembled a combined dataset from above-mentioned six GEO studies, yielding 1,656 quality--controlled samples across ten tissue types (frontal cortex, entorhinal cortex, superior temporal gyrus, cerebellum, whole blood, prefrontal cortex, dorsolateral prefrontal cortex, hippocampus, middle temporal gyrus, and temporal cortex). Diagnostic groups were harmonized across cohorts into Alzheimer’s disease (AD) and cognitively normal controls (CN), resulting in 908 AD cases and 748 controls. In addition, we evaluated model generalizability using an independent DNA methylation dataset from the ADNI, which provides blood--based methylation profiles with harmonized diagnostic categories and detailed clinical metadata. (Data used in the preparation of this article were obtained from the ADNI database (adni.loni.usc.edu). The ADNI was launched in 2003 as a public-private 
partnership, led by Principal Investigator Michael W. Weiner, MD. The primary goal of ADNI has been to 
test whether serial magnetic resonance imaging (MRI), positron emission tomography (PET), other 
biological markers, and clinical and neuropsychological assessment can be combined to measure the 
progression of MCI and early AD.)

\begin{table}[htbp]
\centering
\caption{Participant demographics and clinical characteristics for the combined GEO dataset and the ADNI validation dataset.}
\label{tab:demographics}
\begin{tabular}{lcccc}
\hline
\textbf{Dataset} & \textbf{Sample} & \textbf{\#Subject} & \textbf{Age (mean $\pm$ SD)} & \textbf{Sex (M/F)} \\
\hline
Combined & AD        & 908 &83.81 $\pm$ 8.57 & 333 / 575 \\
             & Control   & 748 & 74.64 $\pm$ 13.01 & 401 / 347 \\
\hline
 ADNI         & AD        & 169 & 73.84 $\pm$7.25 & 105 / 64 \\
             & Control   & 212 & 74.00 $\pm$ 5.97 & 104 / 108 \\
\hline
\end{tabular}
\end{table}

\subsection{Quality control and normalization}
All raw methylation data (IDAT files) are processed using the ChAMP (Chip Analysis Methylation Pipeline) R package \cite{tian2017champ}, which provides a standardized workflow for Illumina 450K and EPIC arrays. The pipeline will be applied consistently across batches and tissues to ensure comparability. First, IDAT files will be imported using the \texttt{champ.load} function, followed by probe filtering to remove: (i) probes with detection $p$-value $>0.01$, (ii) probes with bead count $<3$, (iii) non-CpG probes, (iv) probes overlapping known SNPs, and (v) cross-reactive probes as annotated in \cite{zhou2017comprehensive}. After quality control, remaining probes will be normalized using the BMIQ method to correct type I and type II probe bias.

To combine information across batches and tissues, $\beta$-value matrices from all datasets are aligned to a common set of CpG probes, defined as the intersection across batches. A merged $\beta$-matrix is then be created by concatenating aligned matrices, and a corresponding phenotype table will be assembled. Batch effects are assessed and adjusted using ComBat from the \texttt{sva} package, with study origin treated as a batch variable. 

Covariates such as age, sex, and tissue type are harmonized prior to integration: age is $z$-score normalized, sex is encoded as a binary indicator, and tissue or brain region labels are assigned numeric codes. Outlier samples will be identified based on inter-array correlation and principal component analysis and excluded from downstream modeling.

To reduce dimensionality and retain informative loci, we compute the variance of each CpG site within each tissue and select the top $k=5000$ most variable sites. The modeling feature set is defined as the union of CpGs showing high variance across tissues. This strategy enriches loci with strong inter-individual variation while preserving tissue-specific signals. The resulting processed dataset, thus, provides a harmonized, quality-controlled, and dimensionally reduced representation of DNA methylation suitable for cross-tissue machine learning analysis.

\subsection{Model design}

To address the challenges of modeling DNA methylation data, which comprise hundreds of thousands of CpG sites across heterogeneous brain and blood tissues, we developed a convolution–transformer framework, \textbf{MethConvTransformer}. Unlike traditional differential methylation analysis that evaluates each site independently and relies on stringent $p$-value thresholds, our approach is designed to exploit both local and long-range correlations among CpGs. A convolutional encoder first aggregates short-range dependencies and reduces dimensionality, while a transformer backbone captures higher-order, non-linear relationships across distant sites. Importantly, we introduce a CpG-wise linear projection layer that directly quantifies site-level contributions, providing a natural bridge between classical statistical tests and modern deep learning. This design not only improves predictive performance in AD classification but also yields interpretable feature attributions at both the CpG and tissue levels, enabling cross-tissue comparisons of epigenetic alterations.

Formally, let $x_i \in \mathbb{R}^{P}$ denote the methylation profile of subject $i$ across $P$ CpG sites, $z_i \in \mathbb{R}^{K}$ the covariate vector (e.g., age and sex), and $r_i$ the categorical tissue label. To retain interpretability at the CpG level, we first apply a CpG--wise linear projection that produces margin scores for each site:
\begin{equation}\label{eq:linearproject}
    \bm{h}_i = \bm{W}_{\text{lin}} x_i + \bm{b}_{\text{lin}}, \quad \bm{h}_i \in \mathbb{R}^{P},
\end{equation}
where $W_{\text{lin}}$ assigns a weight to each CpG and $b_{\text{lin}}$ is a bias term. This projection serves as both a compressed input representation and a means to attribute predictive importance directly to individual CpGs.  

The projected signal $h_i$ is then processed by a convolutional encoder $f_{\text{conv}}(\cdot)$ that aggregates local dependencies and downsamples the high--dimensional vector to a token sequence
\begin{equation}\label{eq:tokendim}
    \bm{T}_i = f_{\text{conv}}(\bm{h}_i) \in \mathbb{R}^{L \times d}, \quad L \ll P,
\end{equation}
reducing computational cost while preserving short--range CpG correlations.  

The token sequence is passed through a multi--layer transformer backbone $f_{\text{trans}}(\cdot)$:
\begin{equation}\label{eq:trans}
    \bm{H}_i^{(l+1)} = f_{\text{trans}}^{(l)}(\bm{H}_i^{(l)}), \quad \bm{H}_i^{(0)} = T_i,
\end{equation}
where earlier blocks employ efficient attention mechanisms, the final block retains full attention weights for interpretability. To integrate methylation-derived features with subject metadata, we form the final representation as
\begin{equation}\label{eq:concate}
\bm{u}_i \;=\; \mathrm{Pool}(\bm{H}_i^{(L)}) \;\Vert\; E_{\mathrm{cov}}(\bm{z}_i) \;\Vert\; E_{\mathrm{tissue}}(r_i),
\end{equation}
where $H_i^{(L)}\in\mathbb{R}^{L\times D}$ is the token matrix from the last transformer block, and $\mathrm{Pool}(\cdot)$ denotes mean pooling along the token dimension, producing a $D$-dimensional summary of CpG dependencies. The covariate embedding $E_{\mathrm{cov}}(\bm{z}_i) \in \mathbb{R}^{d_{\text{cov}}}$ projects observed subject-level covariates $\bm{z}_i$ (e.g., age and sex) into a compact latent space, while the tissue embedding $E_{\mathrm{tissue}}(r_i)\in\mathbb{R}^{d_r}$ encodes region identity as a learned fixed effect. The concatenated vector $\bm{u}_i \in \mathbb{R}^{D+d_{\text{cov}}+d_r}$ serves as the joint representation for classification, ensuring that predictions incorporate CpG features, covariates, and tissue context in a unified manner.

The diagnostic prediction is then obtained as
\begin{equation}
    \hat{\bm{y}}_i = \mathrm{softmax}(\bm{W}_c \bm{u}_i + \bm{b}_c),
\end{equation}
where $u_i$ denotes the representation of the $i$-th input, $W_c$ and $b_c$ are the learnable weight matrix and bias vector, respectively, and $\mathrm{softmax}(\cdot)$ maps the logits into a probability distribution over classes.

This architecture integrates CpG--wise linear projection, convolutional downsampling, and transformer attention to jointly address the challenges of dimensionality, heterogeneity, and interpretability in cross--tissue methylation analysis. The linear projection assigns explicit weights to individual CpGs, providing site--level importance scores analogous to differential methylation analysis but without reliance on arbitrary significance thresholds. Convolutional layers then compress the extremely high--dimensional input while preserving local correlations, enabling efficient representation learning. The transformer backbone builds on this compressed sequence to capture long--range, cross--CpG dependencies that may reflect distributed epigenetic dysregulation in Alzheimer’s disease. By concatenating learned covariate and tissue embeddings with the pooled transformer output, the model explicitly accounts for demographic variation and tissue context, thereby supporting unified cross--tissue biomarker discovery. 
\subsection{Interpretability and biomarker identification}

To ensure that model predictions can be related back to biologically meaningful entities, we incorporated multiple attribution mechanisms aligned with the implemented architecture. These mechanisms span single CpGs, local regions, subject-specific contributions, and sequence-level dependencies.

% At the CpG level, we include a linear projection layer that directly assigns a weight to each CpG input. Implemented as a $1\times 1$ convolution, this layer maps the raw methylation profile $x_i\in\mathbb{R}^{1\times P}$ to a margin map, as shown in Eq.\ref{eq:linearproject},
% where the coefficient $W_{lin}$ provides a continuous, interpretable importance score for each CpG sites. Because this operation is linear and aligned to the CpG axis, the weights can be viewed analogously to regression coefficients, indicating which CpGs the network prioritizes during prediction.
At the CpG level, we incorporate a depthwise convolution layer designed exclusively for interpretability rather than model performance evaluation. This linear projection independently assigns a weight to each CpG input, transforming the raw methylation profile $x_i \in \mathbb{R}^{1\times P}$ into a margin map, as formulated in Eq.~\ref{eq:linearproject}. The resulting coefficients $W_{\text{lin}}$ yield continuous and directly interpretable importance scores for individual CpG sites. Owing to the strictly linear nature of this operation along the CpG axis, these weights can be interpreted analogously to regression coefficients, thereby elucidating which CpG loci are prioritized by the network during prediction.

While single-site weights identify individual CpGs, epigenetic regulation often arises from spatially correlated blocks of CpGs. To expose such localized patterns, we employ Grad--CAM++ \cite{chattopadhay2018grad, qu2021ensemble} on selected convolutional feature maps. For a given class $c$ with logit $y^c$ and activations $A^k=(A_1^k,\ldots,A_{L'}^k)$ from channel $k$, let $G_t^k=\partial y^c/\partial A_t^k$ denote the gradient of the class score with respect to activation $A_t^k$. Grad--CAM++ assigns position-specific weights
\begin{equation}\label{eq:gradcamscore}
    \alpha_{k,t}^c \;=\;
\frac{(\bm{G}_t^k)^2}{\,2(\bm{G}_t^k)^2 + \sum_{u=1}^{L'} \bm{A}_u^k (\bm{G}_u^k)^3 + \varepsilon\,},
\end{equation}
where
\begin{equation}\label{eq:gradcamweight}
    \bm{w}_k^c \;=\; \sum_{t=1}^{L'} \alpha_{k,t}^c \,\mathrm{ReLU}(\bm{G}_t^k),
\end{equation}
and constructs the class activation map
\begin{equation}\label{eq:gradcammap}
    \bm{M}^c(t) \;=\; \mathrm{ReLU}\!\Big(\sum_{k=1}^{C} \bm{w}_k^c\, \bm{A}_t^k\Big), \qquad t=1,\ldots,L'.
\end{equation}
Here, $L'$ is the sequence length after convolution/pooling, $C$ the number of channels, and $\bm{A}_t^k$ the activation at channel $k$ and position $t$. The weight $\alpha_{k,t}^c$ highlights locations where both the activation $\bm{A}_t^k$ and its sensitivity $\bm{G}_t^k$ provide consistent positive evidence for class $c$, while the denominator stabilizes the weighting by aggregating higher-order terms across positions within the same channel. The channel weights $\bm{w}_k^c$ then summarize position-wise contributions, and the final map $\bm{M}^c$ emphasizes regions that positively support the prediction. Upsampling $\bm{M}^c$ back to CpG resolution produces a saliency curve over $P$ sites, enabling direct biological interpretation. An important advantage of this approach is its flexibility across representation scales: applying Grad--CAM++ to early convolutional layers emphasizes fine-grained, CpG-level signals close to the input, whereas targeting deeper layers highlights more abstract, region-level patterns that emerge after hierarchical pooling. This multiscale interpretability is particularly well suited to DNA methylation data, where both single-site variation and block-level co-methylation carry biological meaning.

To complement global feature weights with subject-specific explanations, we employ SHAP (SHapley Additive exPlanations). The key idea is to view each CpG site as a "player" in a cooperative game, where the model prediction for a subject corresponds to the game payoff. The contribution of a CpG is then defined as its average marginal effect across all possible feature coalitions, ensuring that attributions are both fair and additive. 

For class $c$, the Shapley value of CpG $j$ for subject $i$ is defined as
\begin{equation}\label{eq:shap}
\phi^{(c)}_{j}(\bm{x}_i)
= \sum_{S \subseteq F \setminus \{j\}} w(S)\,\Delta_S^j v_c(\bm{x}_i),
\end{equation}
where $F$ denotes the CpG feature set. 
Here $w(S)=\tfrac{|S|!(|F|-|S|-1)!}{|F|!}$ is the Shapley weight of coalition $S$, 
and $\Delta_S^j v_c(\bm{x}_i)=v_c(S\cup\{j\};\bm{x}_i)-v_c(S;\bm{x}_i)$ quantifies the marginal contribution of CpG $j$ given $S$.

The value function is defined by marginalizing features not in $S$ against a background set $\mathcal{B}$:
\begin{equation}\label{eq:shapbackground}
v_c(S;\bm{x}_i)=\mathbb{E}_{\bm{x}' \in \mathcal{B}}\!\left[
f_c\!\big(\bm{x}_i|_{S}\,\oplus\,\bm{x}'|_{F\setminus S}\big)
\right],
\end{equation}
where $f_c(\cdot)$ is the model’s class-$c$ softmax output, 
and $\oplus$ denotes the substitution of the missing coordinates of $\bm{x}_i$ by those from a background instance $\bm{x}'$. 
In practice, $\mathcal{B}$ is set to a single reference sample, and region- and covariate-related features are held fixed by the wrapper. 
Thus, $\phi^{(c)}_{j}(\bm{x}_i)$ provides class-specific probability attributions with respect to CpG features only. SHAP allows us to identify, for each subject, which CpGs drive the model toward a given class, thereby providing individualized interpretability beyond global importance scores.

Finally, at the sequence level, the last transformer block produces multi-head self-attention matrices that reveal dependencies among CpG tokens. For head $h$ with query and key matrices $\bm{Q}^{(h)},\bm{K}^{(h)}\in\mathbb{R}^{L\times d_h}$,
\begin{equation}\label{eq:selfatt}
   \bm{A}^{(h)} \;=\; \mathrm{softmax}\!\left(\frac{\bm{Q}^{(h)} \bm{K}^{(h)\top}}{\sqrt{d_h}}\right) \in \mathbb{R}^{L\times L}, 
\end{equation}
and the averaged map across $H$ heads,
\begin{equation}\label{eq:attmap}
    \bar{\bm{A}} \;=\; \tfrac{1}{H}\sum_{h=1}^H \bm{A}^{(h)},
\end{equation}
is retained for interpretation. These matrices quantify how CpG tokens attend to one another within a sample, capturing long-range dependencies introduced by the transformer backbone. Because region embeddings are added only at the classifier head, the attention maps represent pure CpG-token relationships that are independent of tissue or covariates.

Together, these interpretability mechanisms provide a multi-resolution view of model decisions. 
The CpG-wise linear projection connects model parameters directly to individual methylation sites, offering a global effect size interpretation. Grad--CAM++ complements this by highlighting local correlation blocks, which better capture biological realities, such as differentially methylated regions. 
Finally, SHAP provides subject-specific attributions, explaining why a particular sample is assigned to a given class. By combining global site-level weights, regional patterns, and individualized feature contributions, our framework delivers a coherent and biologically grounded interpretability profile that is essential for robust biomarker discovery in high-dimensional methylation data.

\subsection{Model training}
The dataset is split into training, validation, and test sets using stratified sampling ($80/10/10$ by default) to preserve class proportions. Models are optimized with Adam, applying distinct weight decay to the CpG projection layer and the rest of the network. Early stopping is based on validation AUC.  

The training objective combines a label-smoothed cross-entropy loss with a CpG-wise margin regularizer to enhance both predictive robustness and interpretability. Formally, for a mini-batch $\mathcal{B}$ we minimize
\begin{equation}\label{eq:loss}
\mathcal{L} 
= \frac{1}{|\mathcal{B}|}\sum_{i\in\mathcal{B}} 
\Big[ \mathcal{L}_{\text{CE}}(\tilde{\bm{y}}_i,\hat{\bm{y}}_i) 
\;+\; \alpha \,\mathcal{R}_{\text{margin}}(\bm{y}_i^{\pm1},\bm{m}_i) \Big],
\end{equation}
where $\mathcal{L}_{\text{CE}}$ denotes the cross-entropy with label smoothing, $|\mathcal{B}|$ denotes the size of the mini batch $\mathcal{B}$, $\mathcal{R}_{\text{margin}}$ is a CpG-specific margin penalty, and $\mathcal{L}_{\text{CE}}$ is the cross-entropy term defined as \[\mathcal{L}_{\text{CE}}(\tilde{\bm{y}}_i,\hat{\bm{y}}_i)
= -\sum_{c=1}^C \tilde{\bm{y}}_{ic}\,\log \hat{\bm{y}}_{ic},
\]
where $\hat{\bm{y}}_i=\mathrm{softmax}(\bm{z}_i)$ are the  predicted class probabilities from logits $\bm{z}_i$. We further apply label smoothing in the cross-entropy loss to prevent overconfidence and improve generalization. The smoothed target for subject $i$ and class $c$ is defined as
\begin{equation}\label{eq:smooth}
\tilde{\bm{y}}_{ic} \;=\;
\begin{cases}
1 - \varepsilon + \dfrac{\varepsilon}{C}, & \text{if } c = \bm{y}_i, \\[8pt]
\dfrac{\varepsilon}{C}, & \text{if } c \neq \bm{y}_i,
\end{cases}
\end{equation}
where $\bm{y}_i$ is the true class label for sample $i$, 
$c \in \{1, \ldots, C\}$ denotes a candidate class, 
$C$ is the total number of classes, 
and $\varepsilon \in [0,1)$ is the label smoothing factor. When $\varepsilon=0$, the target reduces to the standard one-hot encoding. 

The margin regularizer operates on CpG-wise margin scores $\bm{m}_i=(m_{i,1},\dots,m_{i,P})$, encouraging the model to separate classes at the feature level. It is defined as
\begin{equation}\label{eq:marginloss}
\mathcal{R}_{\text{margin}}(\bm{y}_i^{\pm1},\bm{m}_i)
= \frac{1}{k}\sum_{t \in \text{Top-}k} \max\!\big(0,\,1 - \bm{y}_i^{\pm1} m_{i,t}\big),
\end{equation}
where $\bm{y}_i^{\pm1}=2\bm{y}_i-1 \in \{-1,+1\}$ is the binary label encoding, and the summation is restricted to the $k$ CpGs with the largest hinge residuals $r_{i,t}=\max(0,1-\bm{y}_i^{\pm1} m_{i,t})$. This “hard-site” selection ensures that the loss focuses on the most discriminative or poorly classified CpGs, rather than diluting gradients over all sites. The trade-off parameter $\alpha>0$ controls the strength of this regularization. Cross-entropy with label smoothing stabilizes training and mitigates overfitting, particularly under class imbalance. The CpG-wise margin regularizer provides two additional benefits: (i) it enforces a biologically meaningful margin at the site level, making feature importance more robust; and (ii) the top-$k$ design concentrates supervision on the most challenging CpGs, improving both convergence efficiency and interpretability. 

Hyperparameters such as learning rate, hidden dimension, number of layers, dropout, label smoothing, and weight decay are tuned using Bayesian optimization (Optuna) \cite{akiba2019optuna} with a Tree-structured Parzen Estimator (TPE) sampler and median pruning for early stopping of underperforming trials. Search spaces are defined over discrete and continuous ranges, and the best configuration is selected according to validation AUC. Performance is assessed on the held-out test set using AUC, accuracy, and F1-score. To ensure robustness, each experiment is repeated under 10 different random seeds, and we report the mean and standard deviation of all metrics across runs.

\begin{figure*}[htp!]
  \centering
  \includegraphics[width=\textwidth]{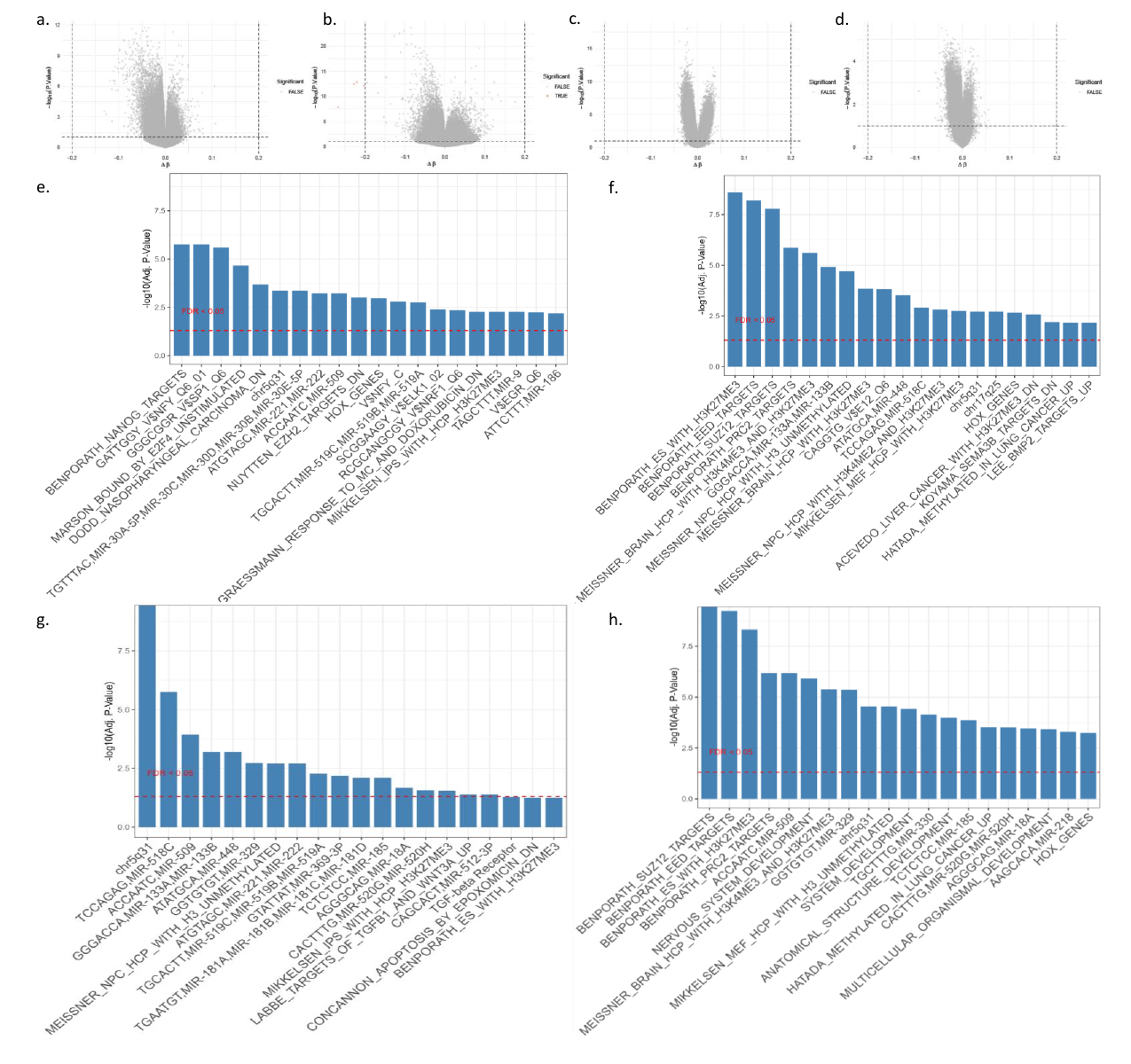}
  \caption{\textbf{Differential methylation profiles across individual datasets.}
(a-d): \texttt{Volcano plots} display sitewise differential methylation (raw $p$ values, $|\Delta\beta|>0.2$) and show that genome-wide significant DMPs are sparse.
(e-h): \texttt{DMR-based gene set enrichment analyses} ($-\log_{10}$ FDR) reveal heterogeneous pathway enrichment across datasets.
% (I-L): \texttt{Density plots} of $\beta$-values illustrate typical bimodal distributions without apparent batch artifacts.
Together, these results indicate that isolated site-level testing yields limited and inconsistent signal, motivating integrative modeling.}
  \label{fig:quad_volcano}
\end{figure*}

\section{Results and Analysis}
\subsection{Limited differential methylation in individual datasets}
We conduct differential methylation analyses on each GEO dataset independently using the ChAMP pipeline, with false discovery rate (FDR) correction. For each dataset, both differentially methylated positions (DMPs) and differentially methylated regions (DMRs) are systematically evaluated, as shown in Fig.\ref{fig:quad_volcano}. Across nearly all datasets, CpG-wise differential testing yields little reproducible signal. In GSE125895 (entorhinal cortex, hippocampus, dorsolateral prefrontal cortex, cerebellum), no sites survive FDR correction, and nominal associations show minimal overlap or consistent direction across regions. GSE134379 (middle temporal gyrus, cerebellum) and GSE80970 (prefrontal, superior temporal cortex) display the same pattern: volcano plots are tightly centered around $\Delta\beta\approx0$ with few outliers, and the top-ranked loci are region- and dataset-specific without cross-regional replication. Sorted neuronal and glial fractions in GSE66351 likewise produce no robust CpG associations after covariate adjustment. In the multi-tissue study GSE59685, a small number of cortical loci approach nominal significance within single tissues but fail to reproduce across tissues, and the matched whole-blood arm shows an essentially null pattern. GSE144858, a whole-blood cohort, also exhibits a symmetric, null-like volcano with no CpGs passing FDR thresholds.

The distributional evidence is consistent with these volcano patterns, as shown in Fig.\ref{fig:quad_volcano} A-D. Groupwise $\beta$-value densities remain bimodal in the expected way and overlap extensively between diagnostic groups, with no global shift or variance change that would indicate broad methylation differences. Together, these observations point to small absolute effect sizes, instability of nominal findings across brain regions, and especially weak signal in blood. Direct sitewise differential analysis therefore provides limited and non-replicable CpG signals in this setting, underscoring the difficulty of deriving peripheral methylation biomarkers for Alzheimer’s disease using univariate tests alone.

DMR analysis, as shown in \mbox{Fig.~\ref{fig:quad_volcano} E--H}, recovers slightly larger sets of candidate regions in some datasets. In \emph{ADNI} whole blood, GSEA of DMRs highlights Polycomb/bivalency programs (e.g., \texttt{BENPORATH\_SUZ12\_TARGETS}, \texttt{BENPORATH\_EED\_TARGETS}, and \texttt{BENPORATH\_PRC2\_TARGETS}) together with generic developmental and cancer-methylation signatures (e.g., \texttt{HATADA\_METHYLATED\_IN\_LUNG\_CANCER\_UP}), indicating peripheral cell-state regulation rather than neuronal pathology. Other blood datasets show the same pattern: \emph{GSE66351} is enriched for hematopoietic/miRNA target modules (e.g., \texttt{miR-223}, \texttt{miR-150}) and broad intracellular programs, while \emph{GSE80970} is dominated by constitutional and imprinting signals (e.g., \texttt{chrXq27} and \texttt{LOPEZ\_MBD\_TARGETS\_IMPRINTED\_AND\_X\_LINKED}). By contrast, brain datasets exhibit neural and synaptic regulatory themes: \emph{GSE125895} shows strong enrichment for neural lineage and bivalent marks (\texttt{MEISSNER\_BRAIN\_HCP}, \texttt{H3K27me3}/\texttt{H3K4me3}) and neuron-linked microRNAs (e.g., \texttt{miR-138}, \texttt{miR-133b}), and \emph{GSE134379} presents a similar mix of PRC2/bivalency with neural/glia-related microRNA modules. These contrasts are consistent with early tau-vulnerable cortical regions undergoing synaptic degeneration and widespread transcriptional reprogramming in brain, whereas peripheral methylation largely reflects systemic immune and inflammatory processes \cite{gasparoni2018dna}. However, these remain highly dataset-specific and show negligible overlap across studies.

Collectively, these findings indicate that conventional univariate differential methylation analyses provide extremely limited signal, with most datasets yielding no significant CpGs and whole-blood datasets performing worst of all. This lack of reproducibility and cross-tissue concordance highlights the intrinsic limitations of direct sitewise testing and motivates integrative modeling strategies designed to uncover subtle, combinatorial methylation signatures relevant to Alzheimer’s disease.

% \begin{figure*}[!t]
%   \centering
%   \textbf{Differential Methylation Signals by Dataset}\\
%   \texttt{Row 1: DMP Volcano (raw $p$; $|\Delta\beta|>0.2$)\quad Row 2: DMR GSEA ($-\log_{10}$ FDR)\quad Row 3: Sample $\beta$ Density}
%   \vspace{0.5em}
%   \includegraphics[width=\textwidth,trim=36 36 36 36,clip]{Figures/quad_volcano_gsea_density_DMRlargest_20251006_214333.pdf}
%   \caption{Across these four datasets, volcano plots show scarce genome-wide DMP signals; DMR enrichments vary by study and lack concordance; density curves reveal standard $\beta$-distributions without obvious batch artifacts at this stage. These patterns reinforce that direct sitewise testing yields limited, non-reproducible signal and motivate integrative modeling.}
%   \label{fig:quad_volcano}
% \end{figure*}

\subsection{Model performance}
\subsubsection{Individual Datasets Reveals Region-Specific Epigenetic Sensitivity}
We evaluate the proposed framework across multiple independent DNA methylation datasets to classify Alzheimer’s disease (AD) versus cognitively normal controls, covering both brain and peripheral tissues. All datasets are processed under a consistent configuration without dataset-specific fine-tuning to ensure comparability and to reflect the intrinsic heterogeneity in biological signal and sample composition. The classification results, as shown in Table.\ref{tab:dataset_performance}, reveal a distinct tissue-dependent pattern. Datasets derived from cortical and hippocampal regions that are highly susceptible to AD pathology achieve the strongest discriminative performance, with AUC values around 0.90–0.97 and mean accuracy above 0.85. These brain regions are characterized by early tau accumulation, synaptic degeneration, and extensive transcriptional reprogramming, all of which are accompanied by pronounced DNA methylation alterations \cite{shireby2022dna}. In contrast, whole-blood datasets such as ADNI and GSE144858 show substantially weaker classification ability, with AUC values of approximately 0.55–0.66. This reduction in performance is consistent with the fact that peripheral methylation largely reflects systemic immune and inflammatory processes rather than direct neuronal pathology \cite{ramakrishnan2024epigenetic}.

Substantial variability in performance is also observed among datasets derived from anatomically similar cortical areas. This variability is likely driven by differences in cell-type composition, as neuron-enriched and glia-enriched samples exhibit distinct methylation profiles, and by technical factors such as measurement platforms and normalization procedures \cite{alves2024unveiling}. Single-cell methylome studies have demonstrated that methylation and chromatin organization vary substantially across neuronal and non-neuronal cell populations \cite{tian2023single}, indicating that bulk tissue measurements represent a composite of multiple, functionally distinct cellular contributions. Although the present analysis does not employ dataset-specific optimization, the observed gradient across regions and tissues provides a clear biological intuition. Methylation changes associated with AD are most prominent in cortical and hippocampal regions undergoing direct neurodegenerative and glial remodeling \cite{gasparoni2018dna}, whereas peripheral tissues contain weaker but complementary systemic information. These results suggest that AD-related epigenetic alterations are distributed unevenly across tissues, with certain regions contributing disproportionately to disease classification. This observation motivates the subsequent combined-dataset experiment, in which all tissue sources are jointly modeled to capture both shared and region-specific methylation signatures and to quantitatively evaluate their relative contributions to AD pathology.

\begin{table*}[htbp!]
\centering
\caption{Performance summary across datasets and tissue regions.}
\renewcommand{\arraystretch}{1.0}  % vertical spacing
\resizebox{0.9\textwidth}{!}{%
\begin{tabular}{l l c c c}
\hline
\textbf{Dataset} & \textbf{Tissue Region} & \textbf{AUC} & \textbf{ACC} & \textbf{F1} \\
\hline
ADNI & Whole blood & $0.55 \pm 0.07$ & $0.62 \pm 0.02$ & $0.37 \pm 0.14$ \\
GSE125895 & ERC; Hippocampus; DLPFC; Cerebellum & $0.95 \pm 0.04$ & $0.91 \pm 0.09$ & $0.79 \pm 0.29$ \\
GSE134379 & Middle temporal gyrus; Cerebellum & $0.62 \pm 0.04$ & $0.63 \pm 0.03$ & $0.69 \pm 0.05$ \\
GSE144858 & Whole blood & $0.66 \pm 0.11$ & $0.69 \pm 0.06$ & $0.58 \pm 0.15$ \\
GSE59685 & ERC; STG; PFC; Cerebellum; Whole blood & $0.95 \pm 0.09$ & $0.94 \pm 0.07$ & $0.96 \pm 0.05$ \\
GSE66351 & Frontal cortex neurons \& glia & $0.74 \pm 0.11$ & $0.78 \pm 0.07$ & $0.84 \pm 0.05$ \\
GSE80970 & Prefrontal cortex; Superior temporal gyrus & $0.90 \pm 0.08$ & $0.85 \pm 0.12$ & $0.86 \pm 0.10$ \\
\hline
\end{tabular}
}% end resizebox
\begin{tablenotes}
\small
\item \textbf{Note.} Each dataset was evaluated over $n=10$ random seeds (mean $\pm$ std).  
The model was trained under a unified configuration and not fine-tuned for each dataset.  
Abbreviations: ERC = entorhinal cortex; DLPFC = dorsolateral prefrontal cortex; STG = superior temporal gyrus; PFC = prefrontal cortex; MTG = middle temporal gyrus.
\end{tablenotes}
\label{tab:dataset_performance}
\end{table*}

\begin{table*}[h!]
\centering
\caption{Comparison between MethConvTransformer and baseline models (mean ± std and Welch's t-test $p$-values).}
\setlength{\tabcolsep}{5pt}
\renewcommand{\arraystretch}{1.0}
\resizebox{0.9\textwidth}{!}{%
\begin{tabular}{lcccccc}
\hline
\textbf{Model} & \textbf{AUC} & \textbf{$p$} & \textbf{ACC} & \textbf{$p$} & \textbf{F1} & \textbf{$p$} \\
\hline
GaussianNB & 0.554 ± 0.010 & 1.78e-29 & 0.536 ± 0.026 & 4.37e-14 & 0.490 ± 0.085 & 8.57e-07 \\
GradientBoosting & 0.741 ± 0.015 & 6.94e-13 & 0.668 ± 0.011 & 2.18e-16 & 0.706 ± 0.011 & 6.36e-16 \\
KNN & 0.651 ± 0.007 & 3.38e-24 & 0.624 ± 0.011 & 2.50e-20 & 0.666 ± 0.021 & 1.02e-11 \\
LDA & 0.781 ± 0.011 & 2.96e-10 & 0.708 ± 0.016 & 4.06e-09 & 0.740 ± 0.012 & 4.83e-11 \\
LinearSVM & 0.829 ± 0.010 & 2.77e-01 & 0.733 ± 0.028 & 1.67e-03 & 0.778 ± 0.019 & 8.19e-03 \\
LogisticRegression\_L1 & 0.769 ± 0.011 & 3.04e-12 & 0.691 ± 0.019 & 1.32e-09 & 0.724 ± 0.023 & 2.03e-07 \\
LogisticRegression\_L2 & 0.825 ± 0.011 & 3.32e-01 & 0.741 ± 0.015 & 1.09e-02 & 0.770 ± 0.013 & 1.24e-02 \\
MLP & 0.498 ± 0.002 & 1.30e-29 & 0.526 ± 0.031 & 1.03e-14 & 0.398 ± 0.213 & 2.84e-06 \\
RandomForest & 0.701 ± 0.011 & 1.55e-20 & 0.640 ± 0.016 & 7.17e-18 & 0.676 ± 0.024 & 1.06e-12 \\
\textbf{MethConvTransformer} & \textbf{0.842 ± 0.021} & -- & \textbf{0.774 ± 0.022} & -- & \textbf{0.803 ± 0.017} & -- \\
\hline
\end{tabular}}
\begin{tablenotes}
\small
\item \textbf{Note.} Values represent mean ± standard deviation across random seeds. Welch's two-sample $t$-test was used to assess the statistical significance of the performance difference between each baseline model and the MethConvTransformer. Bold indicates the highest mean performance per metric. Smaller $p$-values denote stronger evidence that the difference is statistically significant ($p < 0.05$). LinearSVM and LogisticRegression\_L2 show no significant difference from MethConvTransformer in AUC ($p > 0.05$).
\end{tablenotes}
\label{tab:combdataset_performance}
\end{table*}

\subsubsection{Benchmarking on combined methylation }
To quantitatively assess whether integrating multiple tissue sources improves robustness and generalization, we next performed a combined-dataset experiment in which all brain and peripheral methylomes were jointly modeled within a unified feature space. In this setup, samples from distinct cohorts and tissue types were merged after harmonized preprocessing and normalization, allowing the model to learn global patterns of differential methylation while retaining region-specific context. For comparative evaluation, a suite of baseline classifiers—spanning probabilistic (GaussianNB), linear (logistic regression, linear discriminant analysis, and support vector machines), and nonlinear ensemble approaches (Random Forest and Gradient Boosting)—was trained using identical feature representations and cross-validation splits. This design ensures that observed performance differences reflect modeling capacity rather than data imbalance or preprocessing artifacts.

Compared with single-dataset analyses, the combined framework yields substantially higher discrimination between AD and control samples. As summarized in Table~\ref{tab:combdataset_performance}, the MethConvTransformer model achieves the highest mean performance across all evaluation metrics, with an area under the ROC curve (AUC) of \textbf{0.842 ± 0.021}, classification accuracy of \textbf{0.774 ± 0.022}, and F1-score of \textbf{0.803 ± 0.017}. Classical linear baselines such as logistic regression, linear discriminant analysis (LDA), and linear SVM show moderate yet consistent predictive power, whereas tree-based ensemble models (Gradient Boosting and Random Forest) exhibit greater variability and comparatively weaker generalization. According to Welch’s two-sample $t$-test, the MethConvTransformer achieves significantly higher performance than most baseline models ($p < 0.05$). The only exceptions are LinearSVM and LogisticRegression\_L2, where AUC differences do not reach statistical significance, despite significant gains in accuracy and F1-score, highlighting the value of integrative modeling for revealing complex, multi-compartmental epigenetic signatures and improving disease classification performance.

\subsection{Interpretability analyses}
\subsubsection{Dissecting individual CpG relevance through linear and SHAP attributions}
To elucidate how methylation features contribute to disease classification, we applied a dual interpretability strategy that integrates model-intrinsic and post-hoc attribution analyses. At the CpG level, a linear projection layer was incorporated to assign an explicit coefficient weight to each site, enabling direct estimation of its contribution to the model’s decision margin. These coefficients quantify the direction and magnitude of each CpG’s influence but do not imply strong individual effects; rather, they reflect how multiple weak and spatially distributed signals are linearly aggregated within the network. In parallel, gradient-based SHAP values were computed to capture the marginal impact of perturbing each CpG on the output probability, taking into account nonlinear dependencies among loci, tissues, and covariates. The combination of both approaches allows differentiation between structurally encoded feature relevance (from the projection weights) and context-dependent predictive influence (from SHAP), offering a comprehensive characterization of model interpretability.
\begin{figure}[h!]
    \centering
    \includegraphics[width=0.95\linewidth]{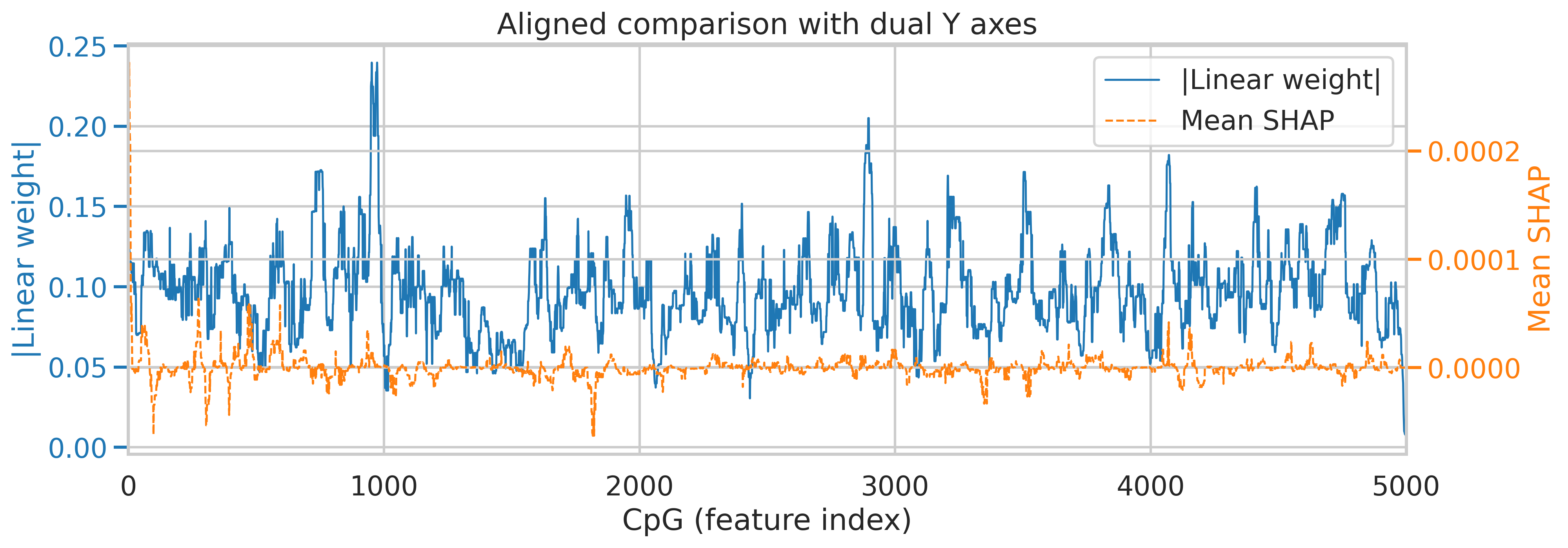}
    \caption{\textbf{Comparison between CpG linear projection weights and SHAP-based feature attributions}. The blue curve shows the absolute coefficients from the linear projection layer, representing intrinsic feature relevance within the network. The orange dashed curve denotes the mean SHAP values across samples, summarizing context-dependent marginal effects. Both profiles reveal that most CpGs exert weak individual effects, and predictive information arises primarily from distributed, cooperative methylation patterns rather than from isolated loci.}
    \label{fig:shaplinear}
\end{figure}
\subsubsection{Inter-site attention patterns and tissue-context modulation}
As shown in Fig.\ref{fig:shaplinear}, the absolute linear weights exhibit high-frequency fluctuations with a few localized peaks, indicating that most CpG sites exert limited individual effects while only a small subset contributes moderately to model discrimination. The mean SHAP profile displays smoother oscillations centered near zero, reflecting the generally weak marginal effects of individual CpGs when averaged across the cohort. This pattern is consistent with results from classical differential methylation analyses in Fig.\ref{fig:quad_volcano}, where few CpGs surpassed statistical thresholds, underscoring that single-site alterations alone are insufficient for robust group separation. Instead, the model integrates numerous subtle and correlated methylation variations to form a predictive representation. Regions showing concordant elevations in both attribution measures denote CpG clusters that contribute consistently across intrinsic and extrinsic interpretability analyses, supporting their potential as cooperative but not dominant predictive loci. At a higher representational level, tissue embeddings were observed to modulate attention maps, emphasizing CpG clusters with elevated inter-tissue methylation variability and thereby anchoring molecular heterogeneity to anatomical context. 
\begin{figure}[h!]
    \centering
    \includegraphics[width=\linewidth,trim={10 10 10 10},clip]{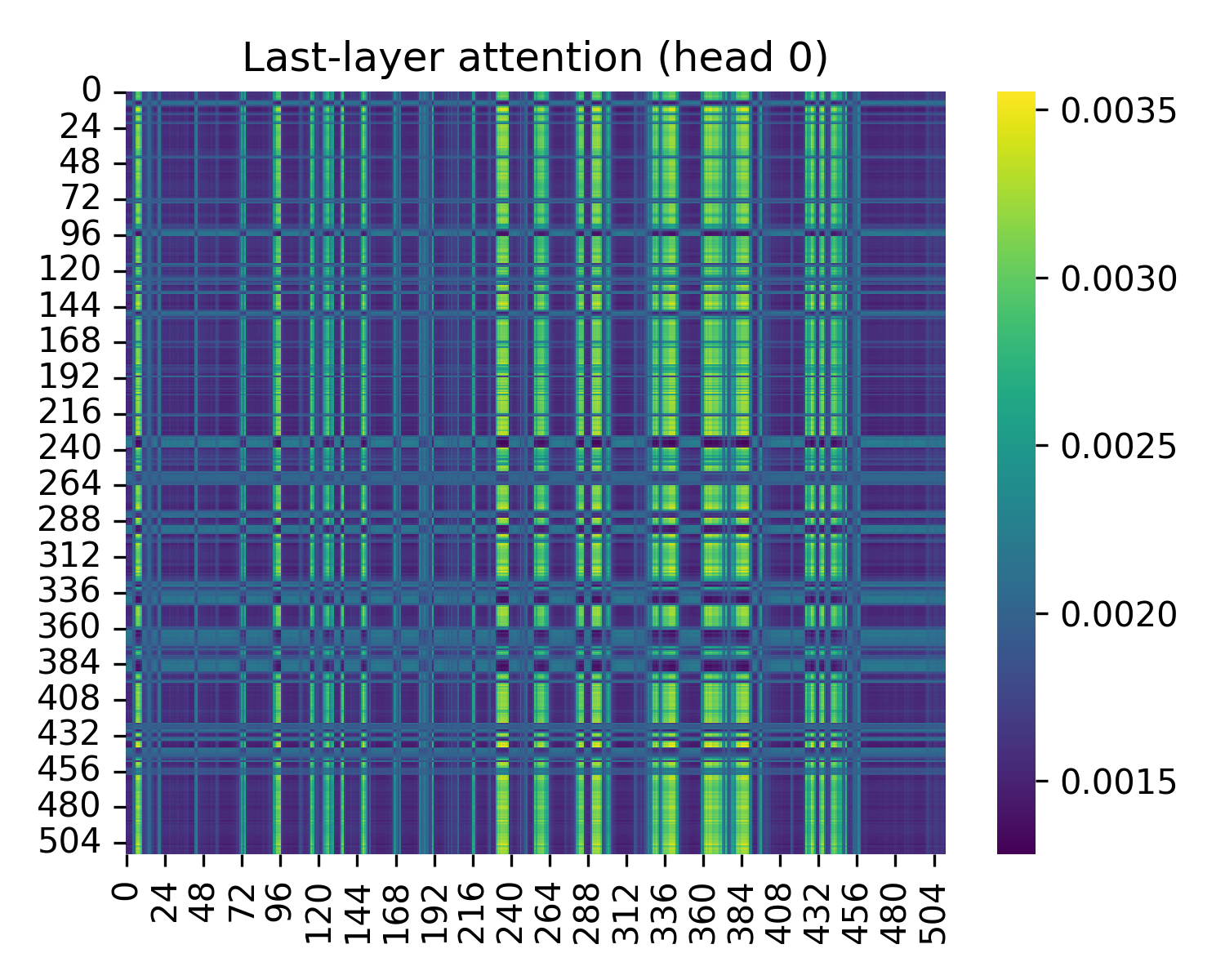}
    \caption{\textbf{Attention pattern in the last Transformer layer.}
    Heatmap of self-attention weights from one representative attention head, illustrating pairwise dependencies among CpG embeddings. Most attention values are near zero, indicating sparse and localized interactions. Periodic cross-shaped hotspots highlight CpG clusters that mutually reinforce each other’s representations, suggesting region-specific co-methylation or shared functional regulation captured by the model.}
    \label{fig:attnmap}
\end{figure}

To further probe how the model integrates information across CpG sites, we examined the attention weights from the last Transformer layer. Figure~\ref{fig:attnmap} illustrates the attention matrix of one representative attention head, showing the pairwise dependency strengths among all input sites. The resulting pattern is characterized by a sparse grid-like structure, where only a limited number of CpG pairs exhibit strong mutual attentions while the majority maintain near-zero weights. This indicates that the model selectively attends to a small subset of interactions rather than uniformly distributing focus across all sites. The observed high-intensity diagonals and periodic cross-links suggest the presence of local, regionally confined attention that may align with known genomic modularity—such as co-regulated CpG clusters or chromatin neighborhood effects—rather than long-range, unstructured dependencies.

Quantitatively, the overall attention entropy remained low across the final layer, implying that attention became more concentrated as the model converged. This focusing behavior suggests that higher layers capture increasingly specific feature relationships that contribute directly to classification. When tissue embeddings were incorporated, the magnitude and spatial spread of these attention hotspots varied systematically across tissues, reflecting how the model contextualizes methylation interactions under different biological environments. Together, these findings support that the model learns biologically meaningful, context-aware co-methylation structures, where only a small subset of CpG interactions are essential for robust prediction.

\begin{figure}[ht!]
    \centering
    \includegraphics[width=\linewidth,trim={10 10 10 10},clip]{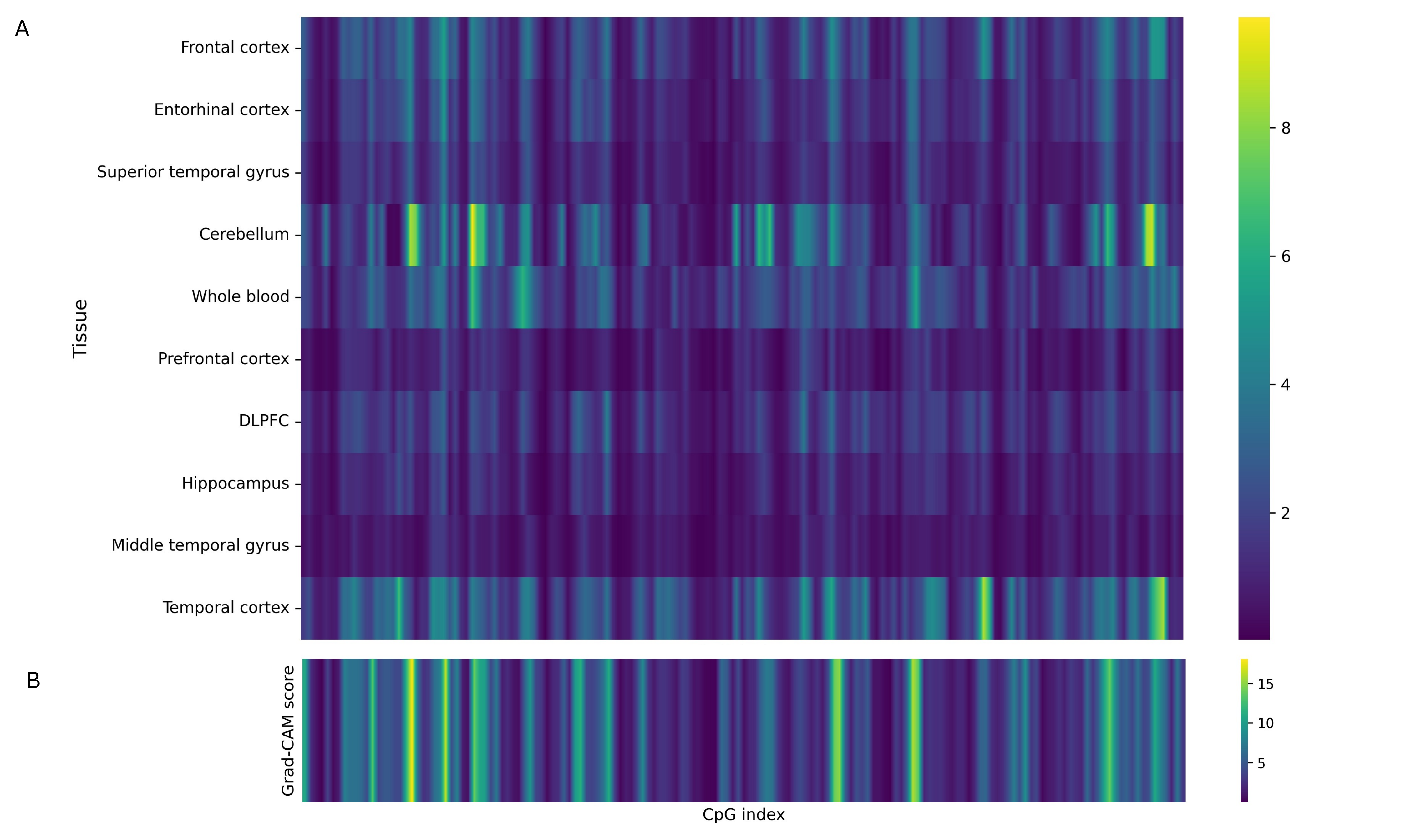}
    \caption{\textbf{Grad-CAM++ interpretability of methylation-based AD classification.}(A) Tissue-specific Grad-CAM++ activation maps showing averaged attribution magnitudes across 10 tissues. Rows correspond to tissues and columns to CpG indices at region level. (B) Overall Grad-CAM++ importance profile aggregated across tissues, highlighting CpG clusters contributing to AD classification.}
    \label{fig:gradcam_tissue}
\end{figure}

\subsubsection{Tissue-specific Grad-CAM++ attributions reveal regional epigenetic vulnerability}

\begin{figure}[!h]
  \centering

  %----------- Subfigure 1 -----------
  \begin{subfigure}[b]{0.5\textwidth}
    \centering
    \includegraphics[width=\textwidth]{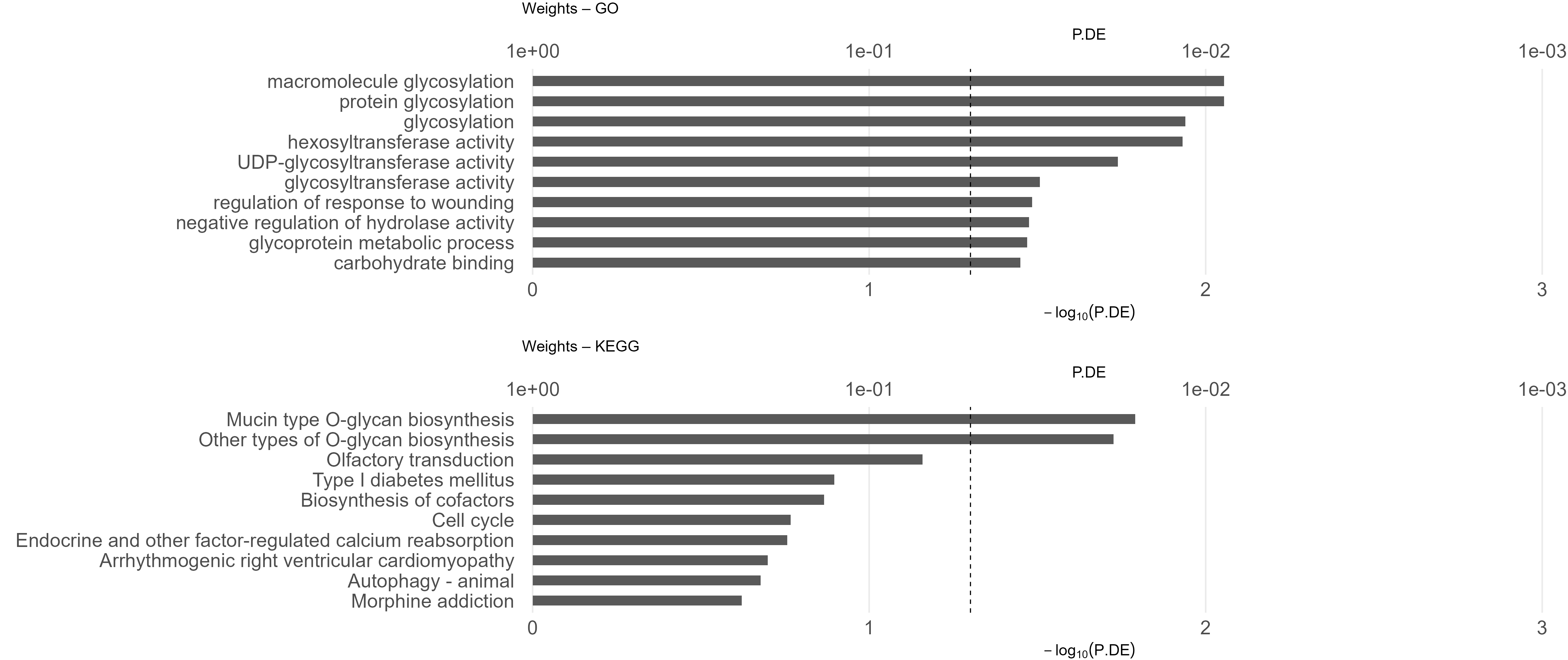}
    \caption{Enrichment using CpGs with the highest absolute linear projection weights, highlighting immune receptor signaling and glycan biosynthesis pathways.}
    \label{fig:enrich_w}
  \end{subfigure}

  \vspace{0em}

  %----------- Subfigure 2 -----------
  \begin{subfigure}[b]{0.5\textwidth}
    \centering
    \includegraphics[width=\textwidth]{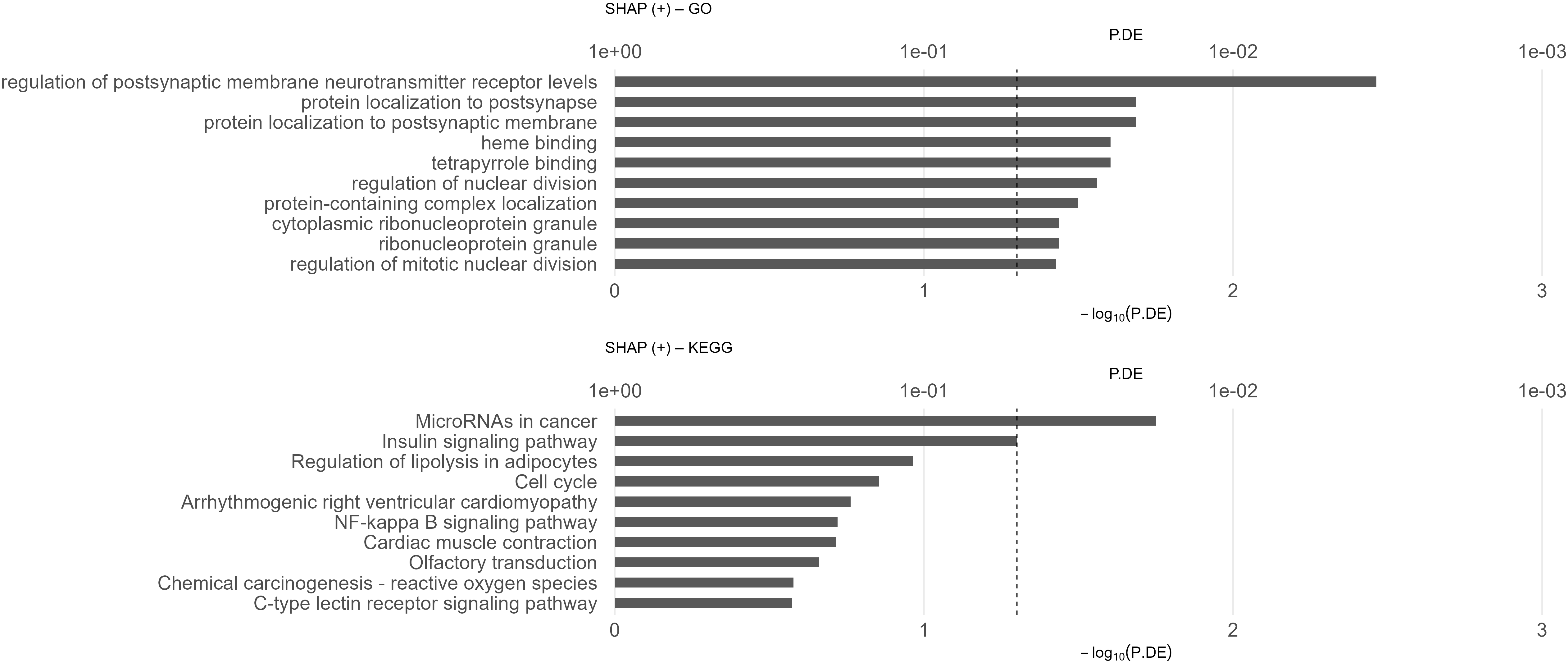}
    \caption{Enrichment using CpGs with the highest positive SHAP values, indicating lipid metabolism, Golgi organization, and energy production processes.}
    \label{fig:enrich_shappos}
  \end{subfigure}

  \vspace{0em}

  %----------- Subfigure 3 -----------
  \begin{subfigure}[b]{0.5\textwidth}
    \centering
    \includegraphics[width=\textwidth]{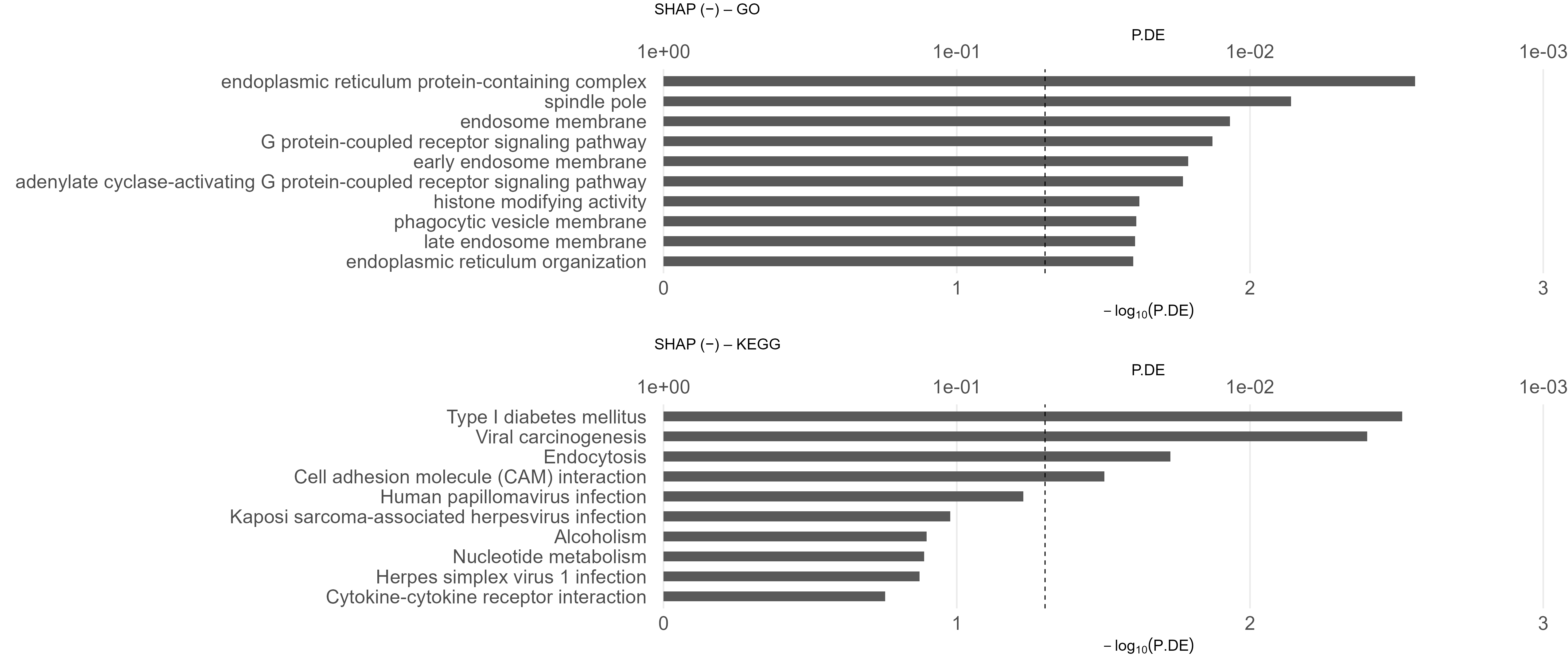}
    \caption{Enrichment using CpGs with the most negative SHAP values, associated with hydrolase activity, GPCR signaling, and endoplasmic reticulum organization.}
    \label{fig:enrich_shapneg}
  \end{subfigure}

  \vspace{0em}

  %----------- Main Caption -----------
  \caption{Gene Ontology (GO) and KEGG pathway enrichment analyses based on CpG sites prioritized by different interpretability layers of the MethConvTransformer model. The enrichment profiles demonstrate that distinct biological processes—ranging from immune activation and glycosylation to lipid metabolism and cellular stress responses—contribute to Alzheimer’s disease–related methylation signatures.}
  \label{fig:enrich}
\end{figure}

The Grad-CAM++ analysis characterizes region-level methylation patterns that drive AD classification (Fig.~\ref{fig:gradcam_tissue}). Unlike SHAP or linear projection analyses, which attribute importance to individual CpG sites, Grad-CAM++ operates on convolutional feature maps and quantifies spatially integrated activation across local CpG neighborhoods. It therefore reflects the collective contribution of regional methylation patterns rather than isolated single-site effects.

The results (Fig.~\ref{fig:gradcam_tissue} A) show that the cerebellum exhibits the strongest Grad-CAM++ activations, followed by the temporal cortex, frontal cortex, entorhinal cortex, and whole blood. All other tissues display comparatively weak activation patterns. The prominence of cerebellar activations indicates that the model assigns high predictive relevance to regional methylation variations in this tissue. This observation is consistent with previous evidence showing that, although the cerebellum is relatively spared from gross neurodegeneration, it undergoes extensive metabolic and epigenetic remodeling in AD \cite{lardenoije2015epigenetics,de2014alzheimer}. The temporal and frontal cortices also demonstrate elevated Grad-CAM++ scores, aligning with regions affected by synaptic loss and cognitive impairment \cite{braak1991neuropathological,arnold1991topographical}, while moderate entorhinal activation corresponds to its established role in early tau pathology. The weaker Grad-CAM++ signals in whole blood suggest reduced discriminative capacity relative to brain tissues but still indicate measurable peripheral reflection of disease-related methylation alterations \cite{smith2021meta,levine2018epigenetic}. It should be noted that sample sizes differ substantially across tissues, which may bias the magnitude of Grad-CAM++ activations, as tissues with more samples contribute proportionally greater gradient information to the averaged maps. Moreover, lower Grad-CAM++ intensity does not imply that CpGs from these tissues are uninformative for AD classification. Rather, it suggests that accurate prediction is difficult when relying on a limited subset of CpGs and that disease-associated methylation patterns are distributed across multiple regions. The overall Grad-CAM++ profile (Fig.~\ref{fig:gradcam_tissue}B) highlights CpG clusters that contribute consistently to classification across tissues, potentially representing shared regulatory modules such as neuroinflammatory and synaptic processes.
\subsubsection{Enrichment Analysis}
We perform Gene Ontology (GO) and KEGG pathway enrichment analyses \cite{ashburner2000gene,kanehisa2000kegg} on CpG sites prioritized by the distinct interpretability layers of our model. We analyze three sets of CpG sites: the top 5\% by absolute model weights, the top 5\% with positive contributions, and the top 5\% with negative contributions. CpGs are mapped to genes using the Illumina HumanMethylation450k annotation \cite{sandoval2011h450k}, from which the background CpGs are selected that test universe matches our feature space. The missMethyl package is apply for enrichment analysis, correcting for unequal CpG probe counts with noncentral hypergeometric (Wallenius) model \cite{phipson2016missmethyl, fog2008calculation}. To account for multiple comparisons, the false discovery rate is controlled using the Benjamini--Hochberg procedure with report raw \(p\) values and FDR \(q\) values reported \cite{benjamini1995fdr}.

Fig.\ref{fig:enrich} summarizes the top biological processes and pathways identified based on weight-derived importance, as well as positive and negative SHAP contributions. Features with high linger projection weight derived importance are predominantly enriched in immune receptor signaling and glycan biosynthesis pathways \cite{haukedal2021implications, kang2024alteration}.  GO terms such as \emph{immune response–activating/response–regulating cell surface receptor signaling}, \emph{protein tyrosine kinase activity}, and \emph{protein autophosphorylation} are highly significant, together with KEGG pathways including \emph{mucin type O–glycan biosynthesis}, \emph{other types of O–glycan biosynthesis}, \emph{cell cycle}, and \emph{endocrine and other factor–regulated calcium reabsorption}.  
These enrichments suggest that our model captures methylation signatures linked to immune regulation and post-translational glycosylation processes \cite{boix2020amyloid, zhang2024glycosylation}. CpGs with positive SHAP contributions are enriched for lipid metabolic processes \cite{yuyama2022linking}, Golgi organization, and cellular energy production.  
Representative GO terms include \emph{glycosphingolipid metabolic process}, \emph{hydrolase activity acting on ester bonds}, \emph{Golgi cisterna}, \emph{Golgi stack}, and \emph{generation of precursor metabolites and energy}, while KEGG pathways such as \emph{microRNAs in cancer}, \emph{insulin signaling}, \emph{regulation of lipolysis in adipocytes}, and \emph{NF–$\kappa$B signaling} highlight metabolic and inflammatory signaling networks.  
These findings indicate that altered lipid handling and endomembrane dynamics represent key components of cross-tissue methylation alterations in Alzheimer’s disease.

Conversely, CpGs with negative SHAP values are associated with hydrolase activity on carbon–nitrogen bonds, G protein–coupled receptor (GPCR) signaling, and endoplasmic reticulum (ER) organization.  
Enriched GO terms include \emph{hydrolase activity acting on carbon–nitrogen (but not peptide) bonds}, \emph{response to ultraviolet light}, \emph{negative regulation of transmembrane transport}, and \emph{endoplasmic reticulum organization}.  
Corresponding KEGG pathways involve \emph{type~I diabetes mellitus}, \emph{viral carcinogenesis}, \emph{endocytosis}, \emph{cell adhesion molecules}, and multiple viral infection pathways (e.g., \emph{human papillomavirus infection} and \emph{herpes simplex virus infection}).  
The presence of metabolic and infectious disease pathways among negatively weighted features suggests that methylation signatures linked to systemic comorbidities \cite{zhang2025dna} may counteract Alzheimer’s disease–related signals, potentially reflecting protective or confounding influences.

Overall, these enrichment results demonstrate that our transformer-based model identifies a broad spectrum of biological processes encompassing immune activation, glycosylation, lipid metabolism, signal transduction, and cellular stress responses.  
The complementary profiles derived from attention weights and SHAP values reveal that attention emphasizes immune and glycosylation mechanisms, whereas SHAP highlights metabolic and signaling processes that modulate disease probability in opposing directions.  
Collectively, these insights enhance the interpretability of the model and generate biologically grounded hypotheses on how cross-tissue DNA methylation contributes to Alzheimer’s disease pathophysiology.

\section{Discussion}
\label{discussion}
\subsection{Key Contributions}
This work introduces \textbf{MethConvTransformer}, a cross‐tissue transformer architecture designed to model both local and long‐range dependencies among CpG sites while explicitly incorporating covariates and tissue context. In benchmarking across six publicly available GEO cohorts and an independent validation set derived from the Alzheimer’s Disease Neuroimaging Initiative (ADNI), the model consistently outperforms classical machine-learning baselines and contemporary deep-learning approaches, demonstrating robustness under heterogeneous preprocessing pipelines, tissue sampling, and cohort composition. These results position MethConvTransformer among the first transformer-based methylation models specifically optimized for \emph{cross-tissue inference} and \emph{multi-resolution interpretability}—two dimensions that constrain most existing epigenetic prediction frameworks.

Compared to the current state of the art in methylation-based disease prediction—which is dominated by elastic-net signatures, tree-based ensemble models, convolutional neural networks (CNNs), and more recently graph- and transformer-based approaches—our method delivers three principal advances. First, we introduce a \emph{CpG-wise linear projection} that aligns deep-learning representations with the effect-size reasoning traditionally employed in epigenetic epidemiology, thereby yielding importance scores that remain interpretable and comparable across tissues. Second, a \emph{convolutional down‐sampling module} preserves short-range co-methylation structure while reducing token length, enabling the transformer attention layers to allocate capacity toward capturing higher‐order dependencies—an architectural advantage over conventional CNN or RNN models. Third, the use of \emph{tissue embeddings and covariate encoders} allows the model to disentangle shared, systemic methylation signatures of Alzheimer’s disease from tissue‐specific or demographic effects without requiring tissue-stratified training. This unified modelling capability directly addresses a major gap in cross‐tissue methylation research, where prior studies largely depend on meta‐analysis or post‐hoc harmonization rather than integrated architectures.

\subsection{Biological Insights and Interpretability Landscape}
Beyond classification performance, MethConvTransformer advances a \emph{multi-level interpretability framework} by integrating global CpG weights, SHAP-derived subject‐specific marginal effects, and Grad-CAM++ attention maps over regional CpG neighbourhoods. These interpretability layers converge on biologically coherent modules. Global weights highlight immune-receptor signalling and glycosylation pathways—processes that have been repeatedly implicated in neuroinflammation, immune–brain crosstalk, and Alzheimer’s disease pathology \cite{zhang2020epigenome}. The SHAP analyses further identify metabolic, vesicular-trafficking and ER/Golgi-organization programmes whose tissue-specific variation modulates class probability—aligning with recent evidence that lipid remodelling, energy metabolism deficits, and endomembrane stress influence neuronal resilience and microglial states. In contrast, negative SHAP features implicate comorbidity‐related programmes (e.g., diabetes and viral response) suggesting that systemic exposures may interfere with the peripheral methylation signatures associated with Alzheimer’s disease.

Taken together, these findings endorse a paradigm increasingly supported by large‐scale epigenome‐wide association studies: Alzheimer’s-related methylation signatures are \emph{polygenic}, \emph{distributed across tissues}, and involve many small‐effect CpGs rather than a few large‐effect loci \cite{smith2021meta}. The ability of MethConvTransformer to recover such distributed patterns across both blood and brain tissues suggests a broader principle: transformer‐based architectures, when structured with interpretable inductive biases such as CpG-wise projection, convolutional down-sampling and tissue embeddings, can capture disease‐relevant biology despite tissue heterogeneity, differing arrays and complex confounding.

\subsection{Clinical and Informatics Implications}
From a translational viewpoint, our results suggest that peripheral methylation signatures alone are unlikely to match the discriminative power of brain-derived methylation profiles (e.g., cortex or hippocampus), but they may nonetheless offer substantial value when integrated in \emph{composite risk models} that combine peripheral methylation with demographic variables, cognitive testing, imaging or fluid biomarkers. The interpretable CpG modules discovered here furnish testable hypotheses for downstream validation, enable the construction of reduced targeted methylation panels (e.g., bisulfite amplicon assays) and enable mechanistic follow-up that is cell-type specific. In practical settings, interpretable cross-tissue methylation models may support early risk stratification, refine patient selection for prevention or early-intervention trials, and facilitate longitudinal monitoring of therapeutic response.

From a biomedical-informatics standpoint, this work demonstrates how high-dimensional epigenetic data (on the order of 100,000+ CpG features) can be modelled using advanced architectures (transformers) while preserving interpretability—an essential prerequisite for clinical deployment. By bridging classical effect-size reasoning with modern deep-learning, and incorporating explicit tissue/context embeddings, MethConvTransformer establishes a blueprint for future high-dimensional biomarker modelling in health informatics.
\subsection{Limitations and future work}
Despite the promising performance and interpretability of the proposed MethConvTransformer framework, several limitations remain. First, the study analyzes retrospective, publicly available cohorts that differ in preprocessing, array platforms, and sampling, which may leave residual batch effects even after harmonization. Second, although the model integrates multiple tissues, the scarcity of matched brain–blood samples constrains direct inference of inter-tissue correspondences and limits causal interpretation. Third, bulk-tissue methylation data average heterogeneous cell populations; incorporating single-cell or sorted-cell methylomes should improve biological resolution and clarify cell-type–specific mechanisms. Fourth, as a data-driven approach, outcomes depend on hyperparameters, optimization dynamics, and initialization, motivating broader robustness probes (e.g., cross-lab replication, perturbation tests, and calibration analysis). Finally, the current work focuses on cross-sectional classification; extending to longitudinal prediction \cite{chen2024explainable} of conversion, rate of decline, or treatment response—and integrating multi-omic or imaging modalities—should enhance translational utility.
Future directions include: (i) prospective validation of candidate CpG panels and pathway modules across sites, (ii) multi-view extensions that jointly encode methylation with transcriptomic, proteomic, or imaging features, (iii) incorporation of explicit cell-type deconvolution or single-cell guidance, and (iv) causal discovery components \cite{wang2024deep}(e.g., invariant risk or counterfactual objectives) to disentangle disease biology from cohort and comorbidity effects.

\section{Conclusion}
\label{conclusion}
We present MethConvTransformer, an interpretable cross-tissue transformer that unifies CpG-wise linear attribution, convolutional context, and self-attention to capture distributed methylation dependencies relevant to Alzheimer’s disease. The model achieves state-of-the-art performance under heterogeneous cohorts and yields stable, biologically coherent attributions that converge on immune signaling, glycosylation, lipid metabolism, and ER/Golgi organization. By reconciling predictive accuracy with multi-resolution interpretability, the framework advances reproducible epigenetic biomarker discovery and provides testable, pathway-level hypotheses. These characteristics position MethConvTransformer as a practical foundation for prospective validation, targeted assay design, and longitudinal monitoring in AD, and as a generalizable template for cross-tissue methylation modeling in other complex diseases.

\section{Fundings}
This work was supported by National Institutes of
Health (NIH) grants [U01AG079847 and R01LM012806
to Z.Z.]. G.Q. received the CPRIT
Postdoctoral Fellowship in the Biomedical
Informatics, Genomics and Translational Cancer Research
Training Program (BIG-TCR) funded by Cancer Prevention
\& Research Institute of Texas (CPRIT RP210045). We thank
UTHealth Cancer Genomics Core for technical support
(CPRIT RP240610).
\section{Acknowledgment}
Data collection and sharing for the Alzheimer's Disease Neuroimaging Initiative (ADNI) is
funded by the National Institute on Aging (National Institutes of Health Grant
U19AG024904). The grantee organization is the Northern California Institute for Research
and Education. In the past, ADNI has also received funding from the National Institute of
Biomedical Imaging and Bioengineering, the Canadian Institutes of Health Research, and
private sector contributions through the Foundation for the National Institutes of Health
(FNIH) including generous contributions from the following: AbbVie, Alzheimer’s Association;
Alzheimer’s Drug Discovery Foundation; Araclon Biotech; BioClinica, Inc.; Biogen; BristolMyers Squibb Company; CereSpir, Inc.; Cogstate; Eisai Inc.; Elan Pharmaceuticals, Inc.; Eli
Lilly and Company; EuroImmun; F. Hoffmann-La Roche Ltd and its affiliated company
Genentech, Inc.; Fujirebio; GE Healthcare; IXICO Ltd.; Janssen Alzheimer Immunotherapy
Research \& Development, LLC.; Johnson \& Johnson Pharmaceutical Research \&
Development LLC.; Lumosity; Lundbeck; Merck \& Co., Inc.; Meso Scale Diagnostics, LLC.;
NeuroRx Research; Neurotrack Technologies; Novartis Pharmaceuticals Corporation; Pfizer
Inc.; Piramal Imaging; Servier; Takeda Pharmaceutical Company; and Transition
Therapeutics.
\section{Competing Interests}
The authors declare no competing interests.

\section{Author Contributions}
G.Q. conceived the study, performed the experiments, analyzed the results, and drafted the manuscript. G.L. assisted in data preprocessing and analysis and contributed to framework development. Z.Z. co-conceived the study, provided guidance on research design, and critically revised the manuscript. All authors read and approved the final version of the manuscript.

\ifCLASSOPTIONcaptionsoff
  \newpage
\fi

\bibliography{Reference.bib}

@article{ADfact2025,
title = {2025 Alzheimer's disease facts and figures},
author={Alzheimer's Association},
journal = {Alzheimer's \& Dementia},
volume = {21},
number = {4},
pages = {e70235},
doi = {https://doi.org/10.1002/alz.70235},
url = {https://alz-journals.onlinelibrary.wiley.com/doi/abs/10.1002/alz.70235},
eprint = {https://alz-journals.onlinelibrary.wiley.com/doi/pdf/10.1002/alz.70235},
year = {2025}
}

@article{xu2025deaths,
  title={Deaths: Final data for 2022},
  author={Xu, Jiaquan and Murphy, Sherry L and Kochanek, Kenneth D and Arias, Elizabeth},
  journal={National Vital Statistics Reports},
  number={4},
  pages={1},
  year={2025}
}

@article{moore2013dna,
  title={DNA methylation and its basic function},
  author={Moore, Lisa D and Le, Thuc and Fan, Guoping},
  journal={Neuropsychopharmacology},
  volume={38},
  number={1},
  pages={23--38},
  year={2013},
  publisher={Nature Publishing Group}
}

@article{horvath2013dna,
  title={DNA methylation age of human tissues and cell types},
  author={Horvath, Steve},
  journal={Genome Biology},
  volume={14},
  number={10},
  pages={3156},
  year={2013},
  publisher={Springer}
}

@article{sarnowski2023multi,
  title={Multi-tissue epigenetic analysis identifies distinct associations underlying insulin resistance and Alzheimer’s disease at CPT1A locus},
  author={Sarnowski, Chlo{\'e} and others},
  journal={Clinical Epigenetics},
  volume={15},
  number={1},
  pages={173},
  year={2023},
  publisher={Springer}
}

@article{lunnon2014methylomic,
  title={Methylomic profiling implicates cortical deregulation of ANK1 in Alzheimer's disease},
  author={Lunnon, Katie and others},
  journal={Nature Neuroscience},
  volume={17},
  number={9},
  pages={1164--1170},
  year={2014},
  publisher={Nature Publishing Group US New York}
}

@article{de2014alzheimer,
  title={Alzheimer's disease: early alterations in brain DNA methylation at ANK1, BIN1, RHBDF2 and other loci},
  author={De Jager, Philip L and others},
  journal={Nature Neuroscience},
  volume={17},
  number={9},
  pages={1156--1163},
  year={2014},
  publisher={Nature Publishing Group}
}

@article{yokoyama2017dna,
  title={DNA methylation alterations in Alzheimer’s disease},
  author={Yokoyama, Amy S and Rutledge, John C and Medici, Valentina},
  journal={Environmental Epigenetics},
  volume={3},
  number={2},
  pages={dvx008},
  year={2017},
  publisher={Oxford University Press}
}

@article{lang2022methylation,
  title={Methylation differences in Alzheimer’s disease neuropathologic change in the aged human brain},
  author={Lang, Anna-Lena and others},
  journal={Acta Neuropathologica Communications},
  volume={10},
  number={1},
  pages={174},
  year={2022},
  publisher={Springer}
}

@article{zhang2020epigenome,
  title={Epigenome-wide meta-analysis of DNA methylation differences in prefrontal cortex implicates the immune processes in Alzheimer’s disease},
  author={Zhang, Lanyu and others},
  journal={Nature Communications},
  volume={11},
  number={1},
  pages={6114},
  year={2020},
  publisher={Nature Publishing Group UK London}
}

@article{shireby2022dna,
  title={DNA methylation signatures of Alzheimer’s disease neuropathology in the cortex are primarily driven by variation in non-neuronal cell-types},
  author={Shireby, Gemma and others},
  journal={Nature Communications},
  volume={13},
  number={1},
  pages={5620},
  year={2022},
  publisher={Nature Publishing Group UK London}
}

@article{gasparoni2018dna,
  title={DNA methylation analysis on purified neurons and glia dissects age and Alzheimer’s disease-specific changes in the human cortex},
  author={Gasparoni, Gilles and others},
  journal={Epigenetics \& chromatin},
  volume={11},
  number={1},
  pages={41},
  year={2018},
  publisher={Springer}
}

@article{smith2024blood,
  title={Blood DNA methylomic signatures associated with CSF biomarkers of Alzheimer's disease in the EMIF-AD study},
  author={Smith, Rebecca G and others},
  journal={Alzheimer's \& Dementia},
  volume={20},
  number={10},
  pages={6722--6739},
  year={2024},
  publisher={Wiley Online Library}
}

@article{fransquet2018blood,
  title={Blood DNA methylation as a potential biomarker of dementia: a systematic review},
  author={Fransquet, Peter D and Lacaze, Paul and Saffery, Richard and McNeil, John and Woods, Robyn and Ryan, Joanne},
  journal={Alzheimer's \& Dementia},
  volume={14},
  number={1},
  pages={81--103},
  year={2018},
  publisher={Elsevier}
}

@article{kaleck2025replication,
  title={Replication of blood DNA methylomic signatures associated with cerebrospinal fluid levels of YKL-40 and NfL biomarkers},
  author={Kaleck, Timo and others},
  journal={Alzheimer's \& Dementia},
  volume={21},
  number={9},
  pages={e70647},
  year={2025},
  publisher={Wiley Online Library}
}

@article{huang2021machine,
  title={A machine learning approach to brain epigenetic analysis reveals kinases associated with Alzheimer’s disease},
  author={Huang, Yanting and  others},
  journal={Nature Communications},
  volume={12},
  number={1},
  pages={4472},
  year={2021},
  publisher={Nature Publishing Group UK London}
}

@article{chen2022multi,
  title={Multi-task deep autoencoder to predict Alzheimer’s disease progression using temporal DNA methylation data in peripheral blood},
  author={Chen, Li and Saykin, Andrew J and Yao, Bing and Zhao, Fengdi and Alzheimer’s Disease Neuroimaging Initiative},
  journal={Computational and Structural Biotechnology Journal},
  volume={20},
  pages={5761--5774},
  year={2022},
  publisher={Elsevier}
}

@article{ren2020identification,
  title={Identification of methylated gene biomarkers in patients with Alzheimer’s disease based on machine learning},
  author={Ren, Jianting and Zhang, Bo and Wei, Dongfeng and Zhang, Zhanjun},
  journal={BioMed Research International},
  volume={2020},
  number={1},
  pages={8348147},
  year={2020},
  publisher={Wiley Online Library}
}

@article{tian2017champ,
  title={ChAMP: updated methylation analysis pipeline for Illumina BeadChips},
  author={Tian, Yuan and others},
  journal={Bioinformatics},
  volume={33},
  number={24},
  pages={3982--3984},
  year={2017},
  publisher={Oxford University Press}
}

@article{zhou2017comprehensive,
  title={Comprehensive characterization, annotation and innovative use of Infinium DNA methylation BeadChip probes},
  author={Zhou, Wanding and Laird, Peter W and Shen, Hui},
  journal={Nucleic Acids Research},
  volume={45},
  number={4},
  pages={e22--e22},
  year={2017},
  publisher={Oxford University Press}
}

@inproceedings{chattopadhay2018grad,
  title={Grad-cam++: Generalized gradient-based visual explanations for deep convolutional networks},
  author={Chattopadhay, Aditya and Sarkar, Anirban and Howlader, Prantik and Balasubramanian, Vineeth N},
  booktitle={2018 IEEE winter conference on applications of computer vision (WACV)},
  pages={839--847},
  year={2018},
  organization={IEEE}
}

@article{c2022cross,
  title={Cross-tissue analysis of blood and brain epigenome-wide association studies in Alzheimer’s disease},
  author={C. Silva, Tiago and others},
  journal={Nature Communications},
  volume={13},
  number={1},
  pages={4852},
  year={2022},
  publisher={Nature Publishing Group UK London}
}

@inproceedings{akiba2019optuna,
  title={Optuna: A next-generation hyperparameter optimization framework},
  author={Akiba, Takuya and Sano, Shotaro and Yanase, Toshihiko and Ohta, Takeru and Koyama, Masanori},
  booktitle={Proceedings of the 25th ACM SIGKDD international conference on knowledge discovery \& data mining},
  pages={2623--2631},
  year={2019}
}

@article{ramakrishnan2024epigenetic,
  title={Epigenetic dysregulation in Alzheimer’s disease peripheral immunity},
  author={Ramakrishnan, Abhirami and others},
  journal={Neuron},
  volume={112},
  number={8},
  pages={1235--1248},
  year={2024},
  publisher={Elsevier}
}

@article{chen2024explainable,
  title={Explainable spatio-temporal graph evolution learning with applications to dynamic brain network analysis during development},
  author={Chen, Longyun and others},
  journal={NeuroImage},
  volume={298},
  pages={120771},
  year={2024},
  publisher={Elsevier}
}

@article{wang2024deep,
  title={A deep dynamic causal learning model to study changes in dynamic effective connectivity during brain development},
  author={Wang, Yingying and others},
  journal={IEEE Transactions on Biomedical Engineering},
  volume={71},
  number={12},
  pages={3390--3401},
  year={2024},
  publisher={IEEE}
}

@article{tian2023single,
  title={Single-cell DNA methylation and 3D genome architecture in the human brain},
  author={Tian, Wei and others},
  journal={Science},
  volume={382},
  number={6667},
  pages={eadf5357},
  year={2023},
  publisher={American Association for the Advancement of Science}
}

@article{alves2024unveiling,
  title={Unveiling DNA methylation in Alzheimer’s disease: a review of array-based human brain studies},
  author={Alves, Victoria Cunha and Carro, Eva and Figueiro-Silva, Joana},
  journal={Neural Regeneration Research},
  volume={19},
  number={11},
  pages={2365--2376},
  year={2024},
  publisher={Medknow}
}

@article{semick2019integrated,
  title={Integrated DNA methylation and gene expression profiling across multiple brain regions implicate novel genes in Alzheimer’s disease},
  author={Semick, Stephen A and others},
  journal={Acta neuropathologica},
  volume={137},
  number={4},
  pages={557--569},
  year={2019},
  publisher={Springer}
}

@article{brokaw2020cell,
  title={Cell death and survival pathways in Alzheimer's disease: an integrative hypothesis testing approach utilizing-omic data sets},
  author={Brokaw, Danielle L and others},
  journal={Neurobiology of aging},
  volume={95},
  pages={15--25},
  year={2020},
  publisher={Elsevier}
}

@article{smith2018elevated,
  title={Elevated DNA methylation across a 48-kb region spanning the HOXA gene cluster is associated with Alzheimer's disease neuropathology},
  author={Smith, Rebecca G and others},
  journal={Alzheimer's \& Dementia},
  volume={14},
  number={12},
  pages={1580--1588},
  year={2018},
  publisher={Wiley Online Library}
}

@article{roubroeks2020epigenome,
  title={An epigenome-wide association study of Alzheimer's disease blood highlights robust DNA hypermethylation in the HOXB6 gene},
  author={Roubroeks, Janou AY and others},
  journal={Neurobiology of Aging},
  volume={95},
  pages={26--45},
  year={2020},
  publisher={Elsevier}
}

@article{lardenoije2015epigenetics,
  title={The epigenetics of aging and neurodegeneration},
  author={Lardenoije, Roy and others},
  journal={Progress in Neurobiology},
  volume={131},
  pages={21--64},
  year={2015},
  publisher={Elsevier}
}

@article{braak1991neuropathological,
  title={Neuropathological stageing of Alzheimer-related changes},
  author={Braak, Heiko and Braak, Eva},
  journal={Acta neuropathologica},
  volume={82},
  number={4},
  pages={239--259},
  year={1991},
  publisher={Springer}
}

@article{arnold1991topographical,
  title={The topographical and neuroanatomical distribution of neurofibrillary tangles and neuritic plaques in the cerebral cortex of patients with Alzheimer's disease},
  author={Arnold, Steven E and Hyman, Bradley T and Flory, Jill and Damasio, Antonio R and Van Hoesen, Gary W},
  journal={Cerebral Cortex},
  volume={1},
  number={1},
  pages={103--116},
  year={1991},
  publisher={Oxford University Press}
}

@article{smith2021meta,
  title={A meta-analysis of epigenome-wide association studies in Alzheimer’s disease highlights novel differentially methylated loci across cortex},
  author={Smith, Rebecca G and others},
  journal={Nature Communications},
  volume={12},
  number={1},
  pages={3517},
  year={2021},
  publisher={Nature Publishing Group UK London}
}

@article{levine2018epigenetic,
  title={An epigenetic biomarker of aging for lifespan and healthspan},
  author={Levine, Morgan E and others},
  journal={Aging (albany NY)},
  volume={10},
  number={4},
  pages={573},
  year={2018}
}

@article{haukedal2021implications,
  title={Implications of glycosylation in Alzheimer’s disease},
  author={Haukedal, Henriette and Freude, Kristine K},
  journal={Frontiers in Neuroscience},
  volume={14},
  pages={625348},
  year={2021},
  publisher={Frontiers Media SA}
}

@article{boix2020amyloid,
  title={Amyloid precursor protein glycosylation is altered in the brain of patients with Alzheimer’s disease},
  author={Boix, Claudia P and Lopez-Font, Inmaculada and Cuchillo-Iba{\~n}ez, Inmaculada and S{\'a}ez-Valero, Javier},
  journal={Alzheimer's Research \& Therapy},
  volume={12},
  number={1},
  pages={96},
  year={2020},
  publisher={Springer}
}

@article{zhang2024glycosylation,
  title={Glycosylation in aging and neurodegenerative diseases: Glycosylation in aging and neurodegeneration},
  author={Zhang, Weilong and Chen, Tian and Zhao, Huijuan and Ren, Shifang},
  journal={Acta Biochimica et Biophysica Sinica},
  volume={56},
  number={8},
  pages={1208},
  year={2024}
}

@article{yuyama2022linking,
  title={Linking glycosphingolipids to Alzheimer’s amyloid-{\ss}: Extracellular vesicles and functional plant materials},
  author={Yuyama, Kohei and Igarashi, Yasuyuki},
  journal={Glycoconjugate Journal},
  volume={39},
  number={5},
  pages={613--618},
  year={2022},
  publisher={Springer}
}

@article{zhang2025dna,
  title={DNA methylation signature of a lifestyle-based resilience index for cognitive health},
  author={Zhang, Wei and others},
  journal={Alzheimer's research \& therapy},
  volume={17},
  number={1},
  pages={88},
  year={2025},
  publisher={Springer}
}

@article{kang2024alteration,
  title={The alteration and role of glycoconjugates in Alzheimer’s disease},
  author={Kang, Yue and Zhang, Qian and Xu, Silu and Yu, Yue},
  journal={Frontiers in Aging Neuroscience},
  volume={16},
  pages={1398641},
  year={2024},
  publisher={Frontiers Media SA}
}

@article{phipson2016missmethyl,
  author  = {Phipson, Belinda and Maksimovic, Jovana and Oshlack, Alicia},
  year    = {2016},
  title   = {missMethyl: An R package for analysing data from Illumina's HumanMethylation450 microarrays},
  journal = {Bioinformatics},
  volume  = {32},
  number  = {2},
  pages   = {286--288},
  doi     = {10.1093/bioinformatics/btv560}
}

@article{ashburner2000gene,
  author  = {The Gene Ontology Consortium},
  year    = {2000},
  title   = {Gene Ontology: Tool for the unification of biology},
  journal = {Nature Genetics},
  volume  = {25},
  number  = {1},
  pages   = {25--29},
  doi     = {10.1038/75556}
}

@article{kanehisa2000kegg,
  author  = {Kanehisa, Minoru and Goto, Susumu},
  year    = {2000},
  title   = {KEGG: Kyoto Encyclopedia of Genes and Genomes},
  journal = {Nucleic Acids Research},
  volume  = {28},
  number  = {1},
  pages   = {27--30},
  doi     = {10.1093/nar/28.1.27}
}

@article{sandoval2011h450k,
  author  = {Sandoval, Juan and others},
  year    = {2011},
  title   = {Validation of a DNA methylation microarray for 450{,}000 CpG sites in the human genome},
  journal = {Epigenetics},
  volume  = {6},
  number  = {6},
  pages   = {692--702},
  doi     = {10.4161/epi.6.6.16196}
}

@article{benjamini1995fdr,
  author  = {Benjamini, Yoav and Hochberg, Yosef},
  year    = {1995},
  title   = {Controlling the false discovery rate: A practical and powerful approach to multiple testing},
  journal = {Journal of the Royal Statistical Society: Series B (Methodological)},
  volume  = {57},
  number  = {1},
  pages   = {289--300},
  doi     = {10.1111/j.2517-6161.1995.tb02031.x}
}

@article{fog2008calculation,
  title={Calculation methods for Wallenius' noncentral hypergeometric distribution},
  author={Fog, Agner},
  journal={Communications in Statistics—Simulation and Computation{\textregistered}},
  volume={37},
  number={2},
  pages={258--273},
  year={2008},
  publisher={Taylor \& Francis}
}

@article{qu2021ensemble,
  title={Ensemble manifold regularized multi-modal graph convolutional network for cognitive ability prediction},
  author={Qu, Gang and others},
  journal={IEEE Transactions on Biomedical Engineering},
  volume={68},
  number={12},
  pages={3564--3573},
  year={2021},
  publisher={IEEE}
}

@article{qu2023interpretable,
  title={Interpretable cognitive ability prediction: A comprehensive gated graph transformer framework for analyzing functional brain networks},
  author={Qu, Gang and others},
  journal={IEEE transactions on medical imaging},
  volume={43},
  number={4},
  pages={1568--1578},
  year={2023},
  publisher={IEEE}
}

\end{document}